\pdfoutput=1

\documentclass[11pt,twoside,a4paper,cmspaper,final,collab]{cms-tdr}
\def\svnVersion{437982P}\def\cmsCernNoTag{CERN-EP-2017-091}\def\cmsCernDate{\today}\def\cmsMessage{Published in Physical Review D as \href{http://dx.doi.org/10.1103/PhysRevD.96.112003}{\doi{10.1103/PhysRevD.96.112003.}}}

\begin{document}\cmsNoteHeader{FSQ-16-004}

\hyphenation{had-ron-i-za-tion}
\hyphenation{cal-or-i-me-ter}
\hyphenation{de-vices}
\RCS$Revision: 437919 $
\RCS$HeadURL: svn+ssh://svn.cern.ch/reps/tdr2/papers/FSQ-16-004/trunk/FSQ-16-004.tex $
\RCS$Id: FSQ-16-004.tex 437919 2017-12-07 08:28:45Z sikler $
\ifthenelse{\boolean{cms@external}}{\providecommand{\cmsLeft}{upper\xspace}}{\providecommand{\cmsLeft}{left\xspace}}
\ifthenelse{\boolean{cms@external}}{\providecommand{\cmsRight}{lower\xspace}}{\providecommand{\cmsRight}{right\xspace}}
\ifthenelse{\boolean{cms@external}}{\providecommand{\cmsLLeft}{Upper\xspace}}{\providecommand{\cmsLLeft}{Left\xspace}}
\ifthenelse{\boolean{cms@external}}{\providecommand{\cmsRRight}{Lower\xspace}}{\providecommand{\cmsRRight}{Right\xspace}}
\ifthenelse{\boolean{cms@external}}{\providecommand{\NA}{\ensuremath{\cdots}\xspace}}{\providecommand{\NA}{\ensuremath{\text{---}}\xspace}}
\newcommand{\mpe}{\ensuremath{\varepsilon}\xspace}
\newcommand{\nh}{\ensuremath{{n_\text{hits}}}\xspace}
\newcommand{\mt}{\ensuremath{m_{\mathrm{T}}}\xspace}
\newcommand{\EPOSLHC}{{\textsc{epos\,lhc}}\xspace}
\newcommand{\EPOS}{\textsc{epos}\xspace}
\newlength\sptw\setlength\sptw{0.07in}

\cmsNoteHeader{FSQ-16-004}

\title{Measurement of charged pion, kaon, and proton production
in proton-proton collisions at \texorpdfstring{$\sqrt{s} =13$\TeV}{sqrt(s) = 13 TeV}}

\date{\today}

\abstract{
Transverse momentum spectra of charged pions, kaons, and protons are
measured in proton-proton collisions at $\sqrt{s} = 13$\TeV with the CMS
detector at the LHC. The particles, identified via their energy loss in the
silicon tracker, are measured in the transverse momentum range of $\pt \approx 0.1$--1.7\GeVc and rapidities $\abs{y} < 1$.
The \pt spectra and integrated yields are compared to previous results
at smaller $\sqrt{s}$ and to predictions of Monte Carlo event generators.
The average \pt increases with particle mass and charged particle
multiplicity of the event. Comparisons with previous CMS results at $\sqrt{s} =0.9$, 2.76, and 7\TeV show
that the average \pt and the ratios of hadron yields feature very similar
dependences on the particle multiplicity in the event, independently of the
center-of-mass energy of the pp collision.}

\hypersetup{%
pdfauthor={CMS Collaboration},%
pdftitle={Measurement of charged pion, kaon, and proton production
in proton-proton collisions at sqrt(s) = 13 TeV},%
pdfsubject={CMS},%
pdfkeywords={CMS, physics, energy loss, hadron spectra}}

\maketitle

\section{Introduction}

\label{sec:introduction}

The study of hadron production has a long history in high-energy particle,
nuclear, and cosmic ray physics.
The absolute yields and the transverse momentum (\pt) spectra of identified
hadrons in high-energy hadron-hadron collisions are among the most basic
physical observables.
They can be used to improve the modeling of various key ingredients of Monte
Carlo (MC) hadronic event generators, such as multiparton interactions, parton
hadronization, and final-state effects (such as parton correlations in color,
\pt, spin, baryon and strangeness number, and collective
flow)~\cite{N.Cartiglia:2015gve}.
The dependence of the hadron spectra and yields on the impact parameter of the
proton-proton (pp) collision provides additional valuable information to tune
the corresponding MC parameters.
Indeed, parton hadronization and final-state effects are mostly constrained
from elementary \Pep\Pem\ collisions, whose final states are largely dominated
by simple \qqbar final states, whereas low-\pt hadrons at the LHC issue from
the fragmentation of multiple gluon ``minijets''~\cite{N.Cartiglia:2015gve}.
Such large differences have a particularly important impact on baryons and
strange hadrons, whose production in pp collisions is not well reproduced by
the existing models~\cite{identifiedSpectra,Adam:2015qaa}, and also affect the
modeling of hadronic interactions of ultrahigh-energy cosmic rays with
Earth's atmosphere~\cite{dEnterria:2011twh}.
Spectra of identified particles in pp collisions also constitute an important
reference for high-energy heavy ion studies, where various final-state effects
are known to modify the spectral shape and yields of different hadron
species~\cite{Khachatryan:2015xaa,Khachatryan:2016odn,Adam:2015kca,Abelev:2012wca,Abelev:2014laa}.

The present analysis uses pp collisions collected by the CMS experiment at the
CERN LHC at $\sqrt{s} = 13$\TeV and focuses on the measurement of the \pt
spectra of charged hadrons, identified primarily via their energy depositions
in the silicon detectors.  The analysis adopts the same methods as used in
previous CMS measurements of pion, kaon, and proton production in pp and pPb
collisions at $\sqrt{s}$ of 0.9, 2.76, and
7\TeV~\cite{identifiedSpectra,Chatrchyan:2013eya}, as well as those performed
by the ALICE Collaboration at 2.76 and 7\TeV~\cite{Aamodt:2011zj,Adam:2015qaa}.

\section{The CMS detector and event generators}

\label{sec:details}

A detailed description of the CMS detector can be found in
Ref.~\cite{:2008zzk}.
The CMS experiment uses a right-handed coordinate system, with the origin at
the nominal interaction point (IP) and the $z$ axis along the
counterclockwise-beam direction.
The pseudorapidity $\eta$ and rapidity $y$ of a particle (in the laboratory
frame) with energy $E$, momentum $p$, and momentum along the $z$ axis $p_z$ are
defined as $\eta = -\ln[\tan(\theta/2)]$, where $\theta$ is the polar angle
with respect to the $z$ axis and $y = \frac{1}{2}\ln[(E+p_z)/(E-p_z)]$.
The central feature of the CMS apparatus is a superconducting solenoid of
6\unit{m} internal diameter. Within the 3.8 T field volume are the silicon
pixel and strip tracker, the crystal electromagnetic calorimeter, and a brass
and scintillator hadron calorimeter.
The tracker measures charged particles within the range $\abs{\eta} < 2.4$. It
has 1440 silicon pixel and 15\,148 silicon strip detector modules with
thicknesses of either 300 or 500\mum, assembled in 13 detection layers in the
central region.
Beam pick-up timing for the experiment (BPTX) devices were used to trigger the
detector readout. They are located around the beam pipe at a distance of
175\unit{m} from the IP on either side, and are designed to provide precise
information on the bunch structure and timing of the incoming beams of the LHC.

In this paper, distributions of identified hadrons produced in inelastic pp
collisions are compared to predictions from MC event generators based
on two different theoretical frameworks: perturbative QCD
({\PYTHIA}6.426~\cite{Sjostrand:2006za} and
{\PYTHIA}8.208~\cite{Sjostrand:2007gs}) and Reggeon field theory
(\EPOS~v3400~\cite{Werner:2005jf}). On the one hand, the basic ingredients of
{\PYTHIA}6 and {\PYTHIA}8 are (multiple) leading-order perturbative QCD $2
\rightarrow 2$ matrix elements, complemented with initial- and final-state
parton radiation (ISR and FSR), folded with parton distribution functions in
the proton, and the Lund string model for parton hadronization. Two different
``tunes'' of the parameters governing the nonperturbative and semihard
dynamics (ISR and FSR showering, multiple parton interactions, beam-remnants,
final-state color-reconnection, and hadronization) are used: the {\PYTHIA}6
Z2*~\cite{Sjostrand:2006za,Khachatryan:2015pea} and {\PYTHIA}8
CUETP8M1~\cite{Khachatryan:2015pea} tunings, based on fits to recent minimum
bias and underlying event measurements at the LHC. On the other hand, {\EPOS}
starts off from the basic quantum field-theory principles of unitarity and
analyticity of scattering amplitudes as implemented in Gribov's Reggeon field
theory~\cite{Gribov:1968fc}, extended to include (multiple) parton scatterings
via ``cut (hard) Pomerons.'' The latter objects correspond to color flux tubes
that are finally hadronized also via the Lund string model. The version of
{\EPOS} used here is run with the LHC tune~\cite{Pierog:2013ria} which includes
collective final-state string interactions resulting in an extra radial flow of
the final hadrons produced in more central pp collisions.

\section{Event selection and reconstruction}

\label{sec:dataAnalysis}

The data used for the measurements presented in this paper were taken during
a special low luminosity run where the average number of pp interactions in
each bunch crossing was 1.0.
A total of $7.0\times 10^6$ collisions were recorded, corresponding to an integrated
luminosity of approximately 0.1\nbinv.

The event selection consisted of the following requirements:

\begin{itemize}

 \item at trigger level, the coincidence of signals from both BPTX devices,
indicating the presence of both proton bunches crossing the interaction point;

 \item offline, to have at least one reconstructed interaction vertex;

 \item beam-halo and beam-induced background events, which usually produce an
anomalously large number of pixel hits, were
identified~\cite{Khachatryan:2010xs} and rejected.

\end{itemize}

The event selection efficiency as well as the tracking and vertexing acceptance
and efficiency are evaluated using simulated event samples produced with the
{\PYTHIA}8 (tune CUETP8M1) MC event generator, followed by the CMS detector
response simulation based on \GEANTfour~\cite{Agostinelli2003250}. Simulated
events are reconstructed and analyzed in the same way as collision data events.
The final results are given for an event selection corresponding to inelastic
pp collisions, which will be presented in Sec.~\ref{sec:results}.
According to the three MC event generators considered, the fraction of inelastic
pp collisions not resulting in a reconstructed pp interaction amounts to about
$14\%\pm3\%$, where the uncertainty is based on the variance of the predictions
coming from the event generators.  These events are mostly diffractive ones
with negligible central activity.

\begin{figure}
 \centering
 \includegraphics[width=0.49\textwidth]{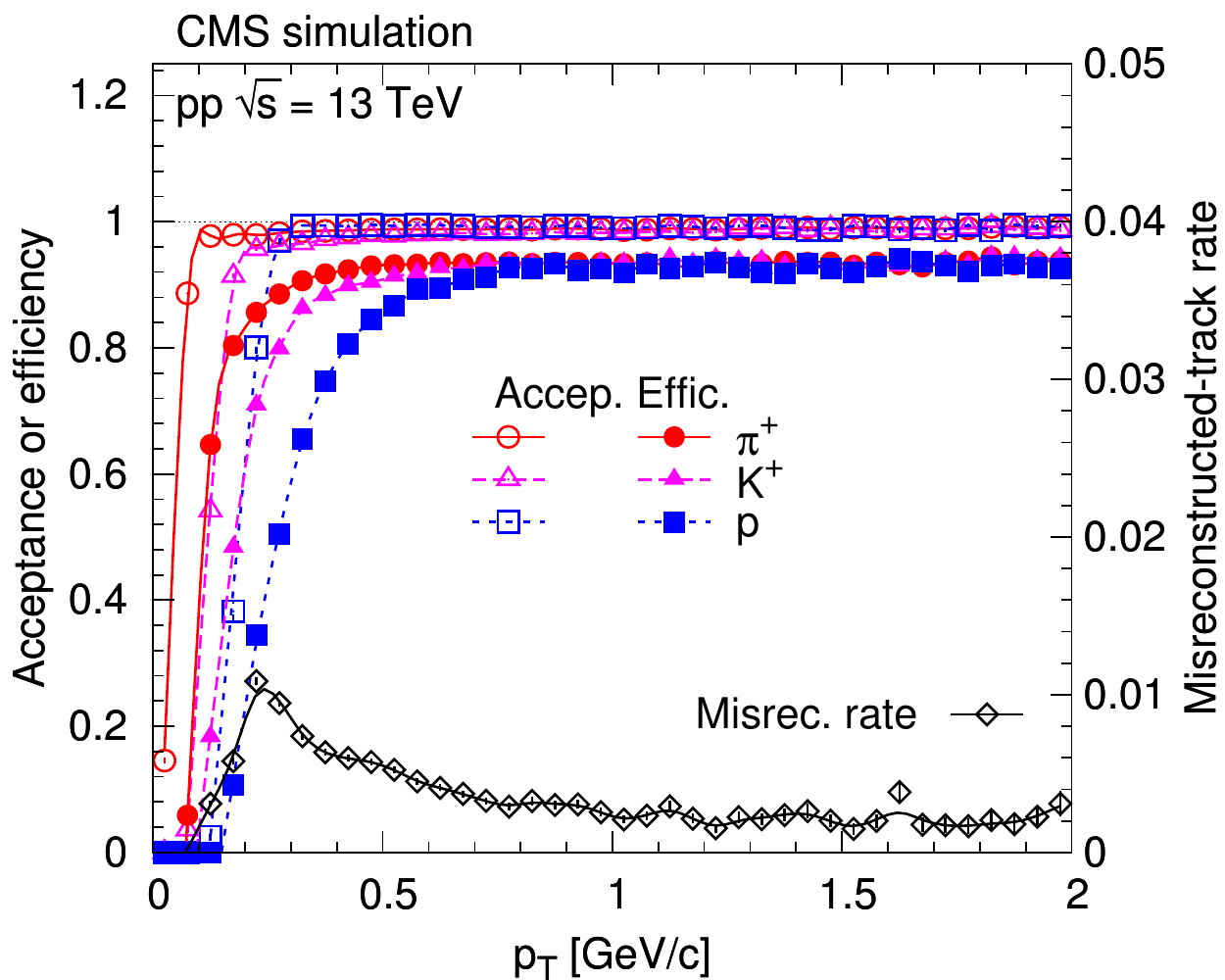}
 \caption{Acceptance (open markers, left scale), tracking efficiency (filled
markers, left scale), and misreconstructed-track rate (right scale) in the
range $\abs{\eta} < 2.4$ as a function of \pt for positively charged pions,
kaons, and protons. The values are very similar for negatively charged
particles.}
 \label{fig:triggerEfficiency}
\end{figure}

The reconstruction of charged particles in CMS is limited by the acceptance of
the tracker ($\abs{\eta} <  2.4$) and by the decreasing tracking efficiency at
low momentum caused by multiple scattering and energy loss.
The identification of particle species using specific ionization
(Sec.~\ref{sec:energyLoss}) is restricted
to $p < 0.15\GeVc$ for electrons, $p < 1.20\GeVc$ for pions, $p < 1.05\GeVc$
for kaons, and $p < 1.70\GeVc$ for
protons~\cite{identifiedSpectra,Chatrchyan:2013eya}.
Pions are measured up to a higher momentum than kaons because of their larger
relative abundance.
In order to have a common kinematic region where pions, kaons, and protons can
all be identified, the range $\abs{y} < 1$ is chosen for this measurement.

The extrapolation of particle spectra into unmeasured $(y,\pt)$ regions is
model dependent, particularly at low \pt. A precise measurement therefore
requires reliable track reconstruction down to the lowest possible \pt
values.
Special tracking algorithms~\cite{Sikler:2007uh}, already used in previous
studies~\cite{Khachatryan:2010xs,Khachatryan:2010us,identifiedSpectra,Chatrchyan:2013eya},
made it possible to extend the present analysis to $\pt \approx 0.1\GeVc$ with
high reconstruction efficiency and low background.
Compared to the standard tracking algorithm used in CMS, these algorithms
feature special track seeding and cleaning, hit cluster shape filtering,
modified trajectory propagation, and track quality requirements.
The charged-pion mass is assumed when fitting particle momenta.

\label{sec:C_x}

The acceptance of the tracker ($C_\mathrm{a}$) is defined as the fraction of
primary charged particles leaving at least two hits in the pixel detector.
Based on MC studies, it is flat in the region $\abs{\eta} < 2$ and $\pt >
0.4\GeVc$, and at values of 96\%--98\% as can be seen in
Fig.~\ref{fig:triggerEfficiency}. The loss of acceptance at $\pt < 0.4\GeVc$ is
caused by energy loss and multiple scattering, which are both functions of
particle mass.
The reconstruction efficiency ($C_\mathrm{e}$), which is defined as the
fraction of accepted charged particles that result in a successfully
reconstructed trajectory, is usually in the range 80\%--90\%.
It decreases at low \pt, also in a mass-dependent way. The
misreconstructed-track rate ($C_\mathrm{f}$), defined as the fraction of
reconstructed primary charged tracks without a corresponding genuine primary
charged particle, is very small, reaching 1\% for $\pt <$ 0.2\GeVc. The
probability of reconstructing multiple tracks ($C_\mathrm{m}$) from a single
charged particle is about 0.1\%, mostly from particles spiralling in the strong
magnetic field of the CMS solenoid. The efficiencies and background rates
(misreconstruction, multiple reconstruction) are found not to depend on the
charged-particle multiplicity of the event in the range of multiplicities of
interest for this analysis. They largely factorize in $\eta$ and \pt, but for
the final corrections (Sec.~\ref{sec:corrections}) an $(\eta,\pt)$ matrix is
used.

The region where pp collisions occur (beam spot) is measured from the
distribution of reconstructed interaction vertices. Since the bunches are very
narrow in the plane transverse to the beam direction (with a width of about
$50\mum$ for this special run), the $x$--$y$ location of the interaction
vertices is well constrained; conversely, their $z$ coordinates are spread over
a relatively long distance and must be determined on an event-by-event basis.
The vertex position is determined using reconstructed tracks that have $\pt >
0.1\GeVc$ and originate from the vicinity of the beam spot, i.e. their
transverse impact parameters $d_\mathrm{T}$ (with respect to the center of the
beam spot) satisfy the condition $d_\mathrm{T} < 3\,\sigma_\mathrm{T}$. Here
$\sigma_\mathrm{T}$ is the quadratic sum of the uncertainty in the value of
$d_\mathrm{T}$ and the root mean square of the beam spot distribution in the
transverse plane.
In order to reach higher efficiency in special-topology low-multiplicity
events, an agglomerative vertex reconstruction algorithm~\cite{Sikler:2009nx}
is used, with the $z$ coordinates of the tracks (and their uncertainties) at
the point of closest approach to the beam axis as input.
The distance distributions of reconstructed vertex pairs in data indicates that
the fraction of merged vertices (with tracks from two or more true vertices)
and split vertices (two or more reconstructed vertices with tracks from a
single true vertex) is about 1\%.
For single-vertex events, there is no minimum requirement on the number of
tracks associated with the vertex (those assigned to it during vertex finding),
and even one-track vertices, which are defined as the point of closest approach
of the track to the beam line, are allowed.
The fraction of events with more than one (three) reconstructed primary
vertices is about 26\% (1.8\%).
Only events with three or fewer reconstructed primary vertices were considered
and only tracks associated with a primary vertex are used in the analysis.

The vertex resolution in the $z$ direction is a strong function of the number
of reconstructed tracks and is always less than 0.1\unit{cm}.
The distribution of the $z$ coordinates of the reconstructed primary vertices
is Gaussian with a width of $\sigma = 4.2\unit{cm}$. Simulated events are
reweighted in order to have the same vertex $z$ coordinate distribution as in
collision data.

The contribution to the hadron spectra from particles of nonprimary origin
arising from the decay of particles with proper lifetime $\tau >
10^{-12}\unit{s}$ was subtracted.
The main sources of these secondary particles are weakly decaying particles,
mostly \PKzS, \PgL/\PagL, and \PgSp/\PagSm. According to the simulations, this
correction ($C_\mathrm{s}$) is approximately 1\% for pions and rises to 15\%
for protons with $\pt \approx 0.2\GeVc$. Because none of these particles decay
weakly into kaons, the correction for kaons is less than 0.1\%.
Charged particles from interactions of primary particles or their decay
products with detector material are suppressed by the impact parameter cuts
described above.

For $p < 0.15\GeVc$, electrons can be clearly identified based on their energy
loss (Fig.~\ref{fig:elossHistos}, left) and their contamination of the hadron
yields is below 0.2\%. Although muons cannot be distinguished from pions,
according to MC predictions their fraction is below 0.05\%. Since both
contaminations are negligible with respect to the final uncertainties, no
corrections are applied.

\section{Estimation of energy loss rate and yield extraction}

\label{sec:energyLoss}

For this paper an analytical parametrization~\cite{Sikler:2011yy} is used to
model the energy loss of charged particles in the silicon detectors. It
provides the probability density $P(\Delta|\varepsilon, l)$ of finding an
energy deposit $\Delta$, if the most probable energy loss rate $\varepsilon$ at
a reference path length $l_0 = 450\mum$ and the path length $l$ are known.
The choice of 450\mum is motivated by being the approximate average path length
traversed in the silicon detectors.
The value of $\varepsilon$ depends on the momentum and mass $m$ of the charged
particle. The parametrization is used in conjunction with a maximum likelihood
fit for the estimate of $\varepsilon$. All details of the methods described
below are given in Ref.~\cite{identifiedSpectra}.

Using the cluster shape filtering mentioned in Sec.~\ref{sec:dataAnalysis},
only hit clusters compatible with the particle trajectory are used.
For clusters in the pixel detector, the energy deposits are calculated based on
the individual pixel deposits.
In the case of clusters in the strip detector, the energy deposits are
corrected for truncation performed by the readout electronics and for losses
due to
deposits below threshold because of capacitive coupling and cross-talk between
neighboring strips. The readout threshold, the strength of coupling, and the
standard deviation of the Gaussian noise for strips are determined from data.
The response of all readout chips is calibrated with multiplicative gain
correction factors.

After the readout chip calibration, the measured energy deposit spectra for
each silicon subdetector are compared to the expectations of the energy loss
model as a function of $p/m$ and $l$ using particles satisfying tight
identification criteria. These comparisons allow the computation of hit-level
corrections to the energy loss model that is used to estimate the particle
energy loss rate $\mpe$ and its associated distribution.

The best value of $\mpe$ for each track is calculated from the measured energy
deposits by minimizing the negative log-likelihood function of the combined
energy deposit for all hits (index $i$) associated with the particle
trajectory, $\chi^2 = -2 \sum_i \ln P(\Delta_i|\varepsilon,l_i)$, where the
probability density functions include the hit-level corrections mentioned
above. Hits with incompatible energy deposits (contributing more than 12 units
to the combined $\chi^2$) are excluded. For the determination of $\mpe$,
removal of at most one hit per track is allowed; this affected about 1.5\% of
the tracks.

\begin{figure}
 \centering
  \includegraphics[width=0.49\textwidth]{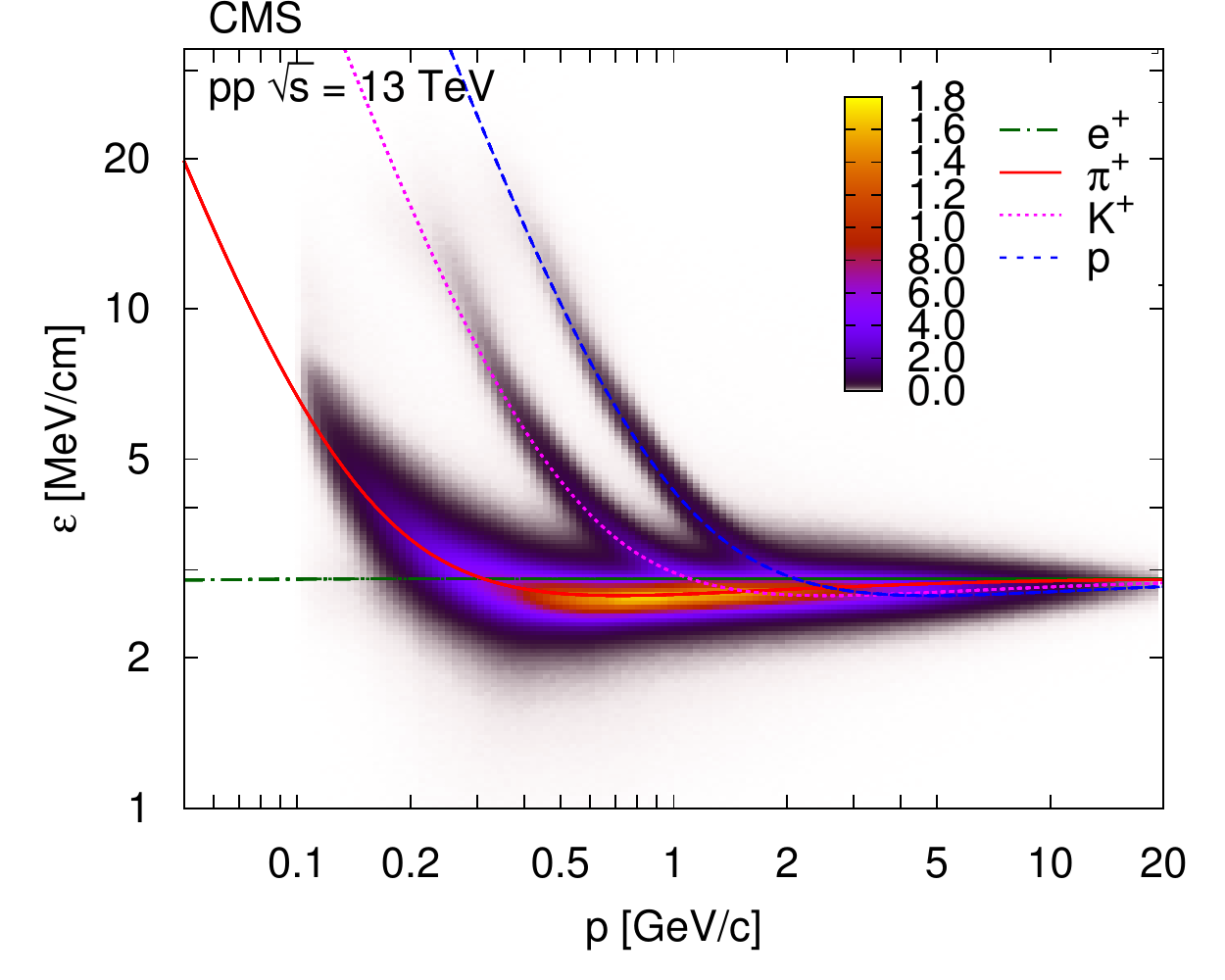}
  \includegraphics[width=0.49\textwidth]{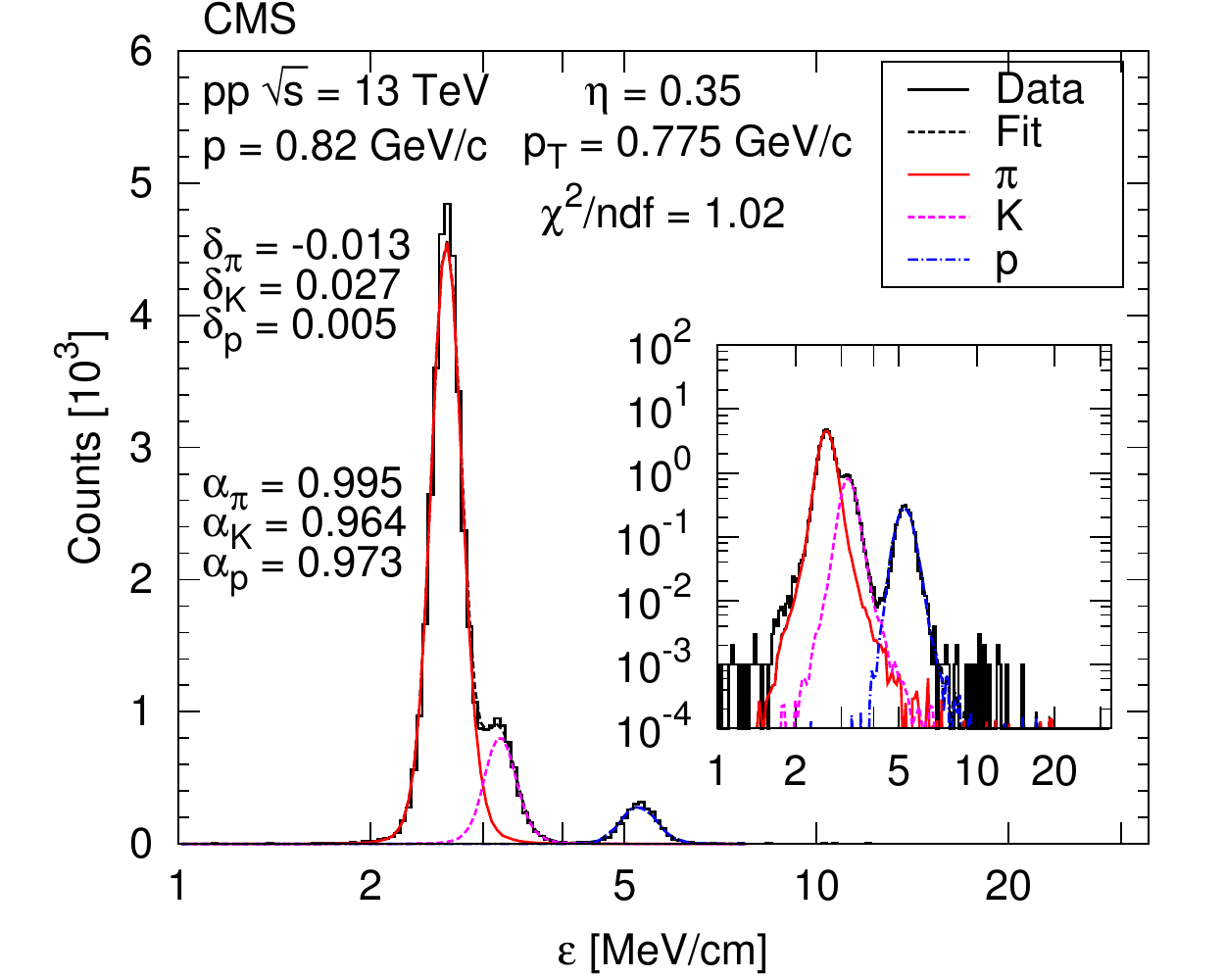}
 \caption{\cmsLLeft: distribution of $\mpe$ as a function of total momentum
$p$, for positively charged reconstructed particles ($\varepsilon$ is the most
probable energy loss rate at a reference path length $l_0 = 450\mum$). The
color scale is shown in arbitrary units and is linear. The curves show the
expected $\mpe$ for electrons, pions, kaons, and protons (Eq.~(30.11) in
Ref.~\cite{Olive:2016xmw}).
\cmsRRight: example $\mpe$ distribution at $\eta = 0.35$ and $\pt = 0.775\GeVc$
(bin centers), with bin widths $\Delta\eta = 0.1$ and $\Delta\pt = 0.05\GeVc$.
Scale factors ($\alpha$) and shifts ($\delta$) are indicated. The inset shows
the distribution with logarithmic vertical scale.}
 \label{fig:elossHistos}
\end{figure}

Low-momentum particles can be identified unambiguously and can therefore be
counted (Fig.~\ref{fig:elossHistos}). Conversely, at high
momenta (above about 0.5\GeVc for pions and kaons and above 1.2\GeVc for
protons) the $\mpe$ bands overlap. Therefore the particle yields need to be
determined by means of a series of template fits in $\mpe$, in bins of $\eta$
and \pt (Fig.~\ref{fig:elossHistos}, \cmsRight panel).
Fit templates with the expected $\mpe$ distributions for all particle species
(electrons, pions, kaons, and protons) are obtained from reconstructed tracks
in data.
All track parameters and hit-related quantities are kept but, in order to
populate the distributions, the energy deposits are regenerated by sampling
from the hit-level corrected analytical parametrization assuming a given
particle type. Possible residual discrepancies between the observed and
expected $\mpe$ distributions, present in some regions of the parameter space
(mostly at low \pt), are taken into account by means of the track-level
corrections consisting, as for the hit-level corrections, of a linear
transformation of the parametrization using scale factors and shifts.
For a less biased determination of these track-level residual corrections,
enriched samples of each particle type are employed for determining starting
values of the parameters to be fitted.
For electrons and positrons, photon conversions in the beam-pipe and in the
innermost pixel layer are used. For high-purity pion and enriched proton
samples, weakly decaying hadrons are selected (\PKzS, \PgL/\PagL).
The following criteria and methods described in
Ref.~\cite{identifiedSpectra} are also exploited to better constrain the
parameters of the fits:
 fitting the $\mpe$ distributions in slices of number of hits ($\nh$) and track
fit $\chi^2/\mathrm{ndf}$ (where ndf is number of degrees of freedom)
simultaneously;
 setting constraints on the $\nh$ distribution for specific particle species;
imposing the expected continuity of track-level residual corrections in
adjacent $(\eta,\pt)$ bins;
 and using the expected convergence of track-level residual corrections as the
$\mpe$ values of two particle species approach each other at large momentum.

Distributions of $\mpe$ as a function of total momentum $p$ for positive
particles are plotted in the \cmsLeft panel of Fig.~\ref{fig:elossHistos} and
compared to the predictions of the energy loss
parametrization~\cite{Sikler:2011yy} for electrons, pions, kaons, and protons.
The results of the (iterative) $\mpe$ fits are the yields for each particle
species and charge in bins of $(\eta,\pt)$ or $(y,\pt)$, both inclusive and
divided into classes of reconstructed primary charged-track multiplicity.
Although pion and kaon yields could not be determined for $p > 1.30\GeVc$,
their sum is measured. This information is an important constraint when fitting
the \pt spectra.

\section{Yield extraction and systematic uncertainties}

\label{sec:corrections}

The measured yields in each $(\eta,\pt)$ bin, $\Delta N_\text{measured}$, are
first corrected for the misreconstructed-track rate $C_\mathrm{f}$ and the
fraction of secondary particles $C_\mathrm{s}$:

\begin{equation}
 \Delta N' = \Delta N_\text{measured}
  \, (1 - C_\mathrm{f}) \, (1 - C_\mathrm{s}).
\end{equation}

The bin widths are $\Delta\eta = 0.1$ and $\Delta\pt = 0.05\GeVc$. The
distributions are then unfolded to take into account bin migrations due to the
finite $\eta$ and \pt resolutions. The $\eta$ distribution of the tracks is
almost flat and the $\eta$ resolution is significantly smaller than the bin
width. At the same time the \pt distribution is steep in the low-momentum
region and separate \pt-dependent corrections in each $\eta$ slice are
necessary.
For that, an unfolding procedure with a linear regularization method (Tikhonov
regularization~\cite{Tikhonov:1963}) is used, based on response matrices
obtained from {\PYTHIA}8 MC samples separately for each particle species.
This procedure guarantees that the uncertainties associated with the assumption
of the pion mass in the track fitting step are taken into account.
The bin purities of the matrices are above 80\%--90\%. The chosen regularization
term reflects that the original distribution changes only slowly, but that that
particular choice has negligible influence on the results.

Further corrections for acceptance, efficiency, and multiple track
reconstruction probability are applied:

\begin{equation}
 \frac{1}{N_\text{ev}} \frac{\rd^2 N}{\rd\eta\, \rd\pt}_\text{corrected} =
  \frac{1}
       {C_\mathrm{a} \, C_\mathrm{e} \, (1 + C_\mathrm{m})}
       \frac{\Delta N'}{N_\text{ev} \Delta\eta \Delta\pt},
\end{equation}

\noindent where $N_\text{ev}$ is the corrected number of inelastic pp
collisions in the data sample.
Bins that meet at least one of the following criteria are not used in order to
ensure robustness of the fits described below and to minimize the impact on the
systematic uncertainties: acceptance less than 50\%; efficiency less than 50\%;
multiple-track rate greater than 10\%; multiplicity below 80 tracks.

Finally, the $\eta$-differential yields $\rd^2N/\rd\eta\,\rd\pt$ are
transformed into $\rd^2N/\rd y\,\rd \pt$ yields by multiplying with the
Jacobian of the $\eta$ to $y$ transformation ($E/p$), and the $(\eta,\pt)$ bins
are mapped onto a $(y,\pt)$ grid.
The differential yields exhibit a slight (5\%--10\%) dependence on $y$ in the
narrow region considered ($\abs{y} < 1$), an effect that decreases with the
event multiplicity.
The yields as a function of \pt are obtained averaged over the rapidity
window.

The \pt distributions are fit using a Tsallis-Pareto-type function, which
empirically describes both the low-\pt exponential and the high-\pt
power-law behaviors while employing only a few parameters. Based on the good
reproduction of previous measurements of unidentified and identified particle spectra
\cite{Khachatryan:2010xs,Khachatryan:2011tm,identifiedSpectra,Chatrchyan:2013eya},
the following form of the distribution \cite{Tsallis:1987eu,Biro:2008hz} is
used:

\begin{gather}
 \frac{\rd^2 N}{\rd y \, \rd\pt} =
  \frac{\rd N}{\rd y} \, C
                \, \pt \left[1 + \frac{\mt - m\,c}{nT} \right]^{-n},
 \label{eq:tsallis}
\intertext{where}
 C = \frac{(n-1)(n-2)}{nT[nT + (n-2) m\,c]}
\end{gather}

\noindent and $\mt = \sqrt{(mc)^2 + \pt^2}$.
The free parameters are the integrated yield $\rd N/\rd y$, the exponent $n$,
and the parameter $T$.
According to some models of particle production based on nonextensive
thermodynamics~\cite{Biro:2008hz}, the parameter $T$ is connected with the
average particle energy, while $n$ characterizes the ``nonextensivity'' of the
process, i.e. the departure of the spectra from a Boltzmann distribution ($n =
\infty$).
Equation~\eqref{eq:tsallis} is useful for extrapolating the spectra down to
zero and up to
high \pt, and thereby extracting $\langle\pt\rangle$ and $\langle\rd
N/\rd y\rangle$. Its validity for different multiplicity bins is cross-checked
by fitting MC spectra in the \pt ranges where there are data points, and
verifying that the fitted values of $\langle\pt\rangle$ and $\langle\rd N/\rd
y\rangle$ are consistent with the generated values.
Nevertheless, for a more robust estimation of both $\langle\pt\rangle$ and
$\langle \rd N/\rd y \rangle$, the unfolded bin-by-bin yield values and their
uncertainties are used in the measured range while the fitted functions are
employed for the extrapolation into the unmeasured regions.

As discussed earlier, pions and kaons cannot be unambiguously distinguished at
high momenta. For this reason the pion-only, the kaon-only, and the joint pion
and kaon $\rd^2N/\rd y \, \rd\pt$ distributions are fitted for $\abs{y} < 1$
and $p < 1.20\GeVc$, $\abs{y} < 1$ and $p < 1.05\GeVc$, and $\abs{\eta} <1$ and
$1.05 < p < 1.7\GeVc$, respectively.
Since the ratio $p/E$ for the pions (which are more abundant than kaons) at
these momenta can be approximated by $\pt/\mt$ at $\eta \approx 0$,
Eq.~\eqref{eq:tsallis} becomes:

\begin{equation}
 \frac{\rd^2 N}{\rd\eta\, \rd\pt} \approx
  \frac{\rd N}{\rd y} \, C
                \, \frac{\pt^2}{\mt} \left(1 + \frac{\mt - m\,c}{nT}
\right)^{-n}.
 \label{eq:tsallis2}
\end{equation}

Moreover, below $\pt$ values of 0.1--0.3\GeV the detector acceptance and
the tracking efficiency significantly decrease. The Tsallis-Pareto function is
used to extrapolate the measured yields both into this latter region and to the
region at high momenta such that the integrated yield ($\rd N/\rd y$) and the
average transverse momentum ($\langle\pt\rangle$) can be reported for the full
$\pt$ range. This choice allows measurements performed by different experiments
in various collision systems and center-of-mass energies to be compared.

The fractions of particles outside the measured \pt range are 15\%--30\% for
pions, 40\%--50\% for kaons, and 20\%--35\% for protons, depending on the track
multiplicity of the event.

\begin{figure}
\centering
  \includegraphics[width=0.49\textwidth]{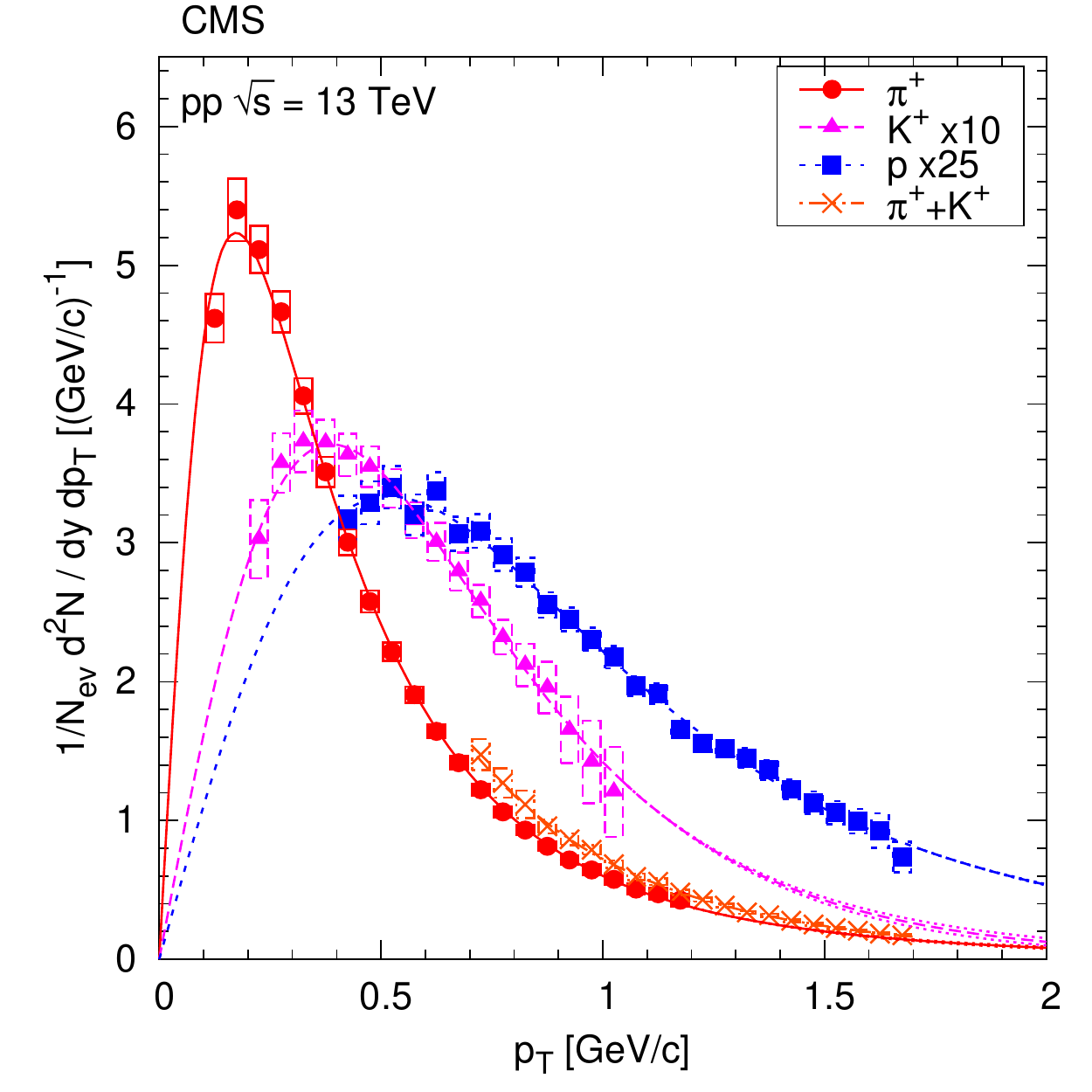}
  \includegraphics[width=0.49\textwidth]{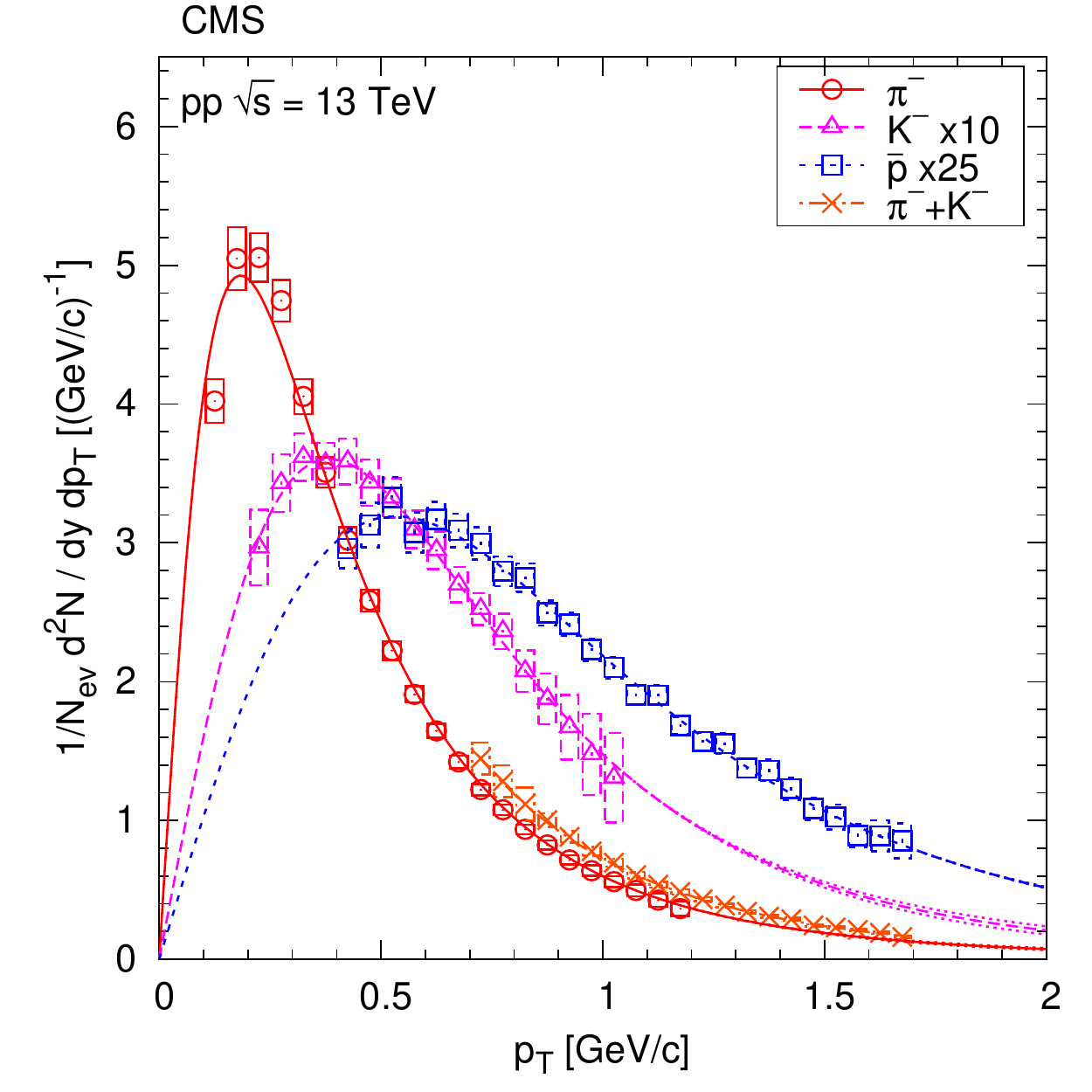}
 \caption{Transverse momentum distributions of identified charged hadrons
(pions, kaons, protons, sum of pions and kaons) from inelastic pp collisions,
in the range $\abs{y}<1$, for positively (\cmsLeft) and negatively (\cmsRight)
charged particles. Kaon and proton distributions are scaled as shown in the
legends. Fits to Eqs.~\eqref{eq:tsallis} and \eqref{eq:tsallis2} are
superimposed.
For the $\Pgp$+$\PK$ fit, only the region corresponding to the range $\abs{\eta} <
1$ and $1.05 < p < 1.7\GeVc$ is plotted.
Boxes show the uncorrelated systematic uncertainties, while error bars indicate
the uncorrelated statistical uncertainties (barely visible).
The fully correlated normalization uncertainty (not shown) is 3.0\%.
Dotted lines (mostly indistinguishable from the nominal fit curves) illustrate
the effect of varying the inverse exponent ($1/n$) of the Tsallis-Pareto
function by $\pm$0.05 beyond the highest-\pt measured point.}
 \label{fig:dndpt_lin}
\end{figure}

The systematic uncertainties are very similar to those in
Ref.~\cite{identifiedSpectra} and are summarized in Table~\ref{tab:error}.
They are obtained from the comparison of different MC event generators,
differences between data and simulation, or based on previous studies (hit
inefficiency, misalignment).
The uncertainties in the corrections $C_\mathrm{a}$, $C_\mathrm{e}$,
$C_\mathrm{f}$ and $C_\mathrm{m}$, which are related to the event selection,
and the effects of pileup, are fully or mostly
correlated and are treated as normalization uncertainties: altogether they
propagate to a 3.0\% uncertainty in the yields and a 1.0\% uncertainty in the
average \pt.
In order to study the influence of the high-\pt extrapolation on the $\langle
\rd N/\rd y\rangle$ and $\langle \pt \rangle$ averages, the reciprocal of the
exponent ($1/n$) of the fitted Tsallis-Pareto function was increased and
decreased by $\pm$0.05 only in the region above the highest measured \pt; in
this same region both the function and its first derivative were required to
fit continuously the data points.
The choice of the magnitude for the variation is motivated by the fitted $1/n$
values and their distance from a Boltzmann distribution. The resulting
functions are plotted in Fig.~\ref{fig:dndpt_lin} as dotted lines (though they
are mostly indistinguishable from the nominal fit curves). The high-\pt
extrapolation introduces systematic uncertainties of 1--3\% for $\langle \rd
N/\rd y\rangle$, and 4--8\% for $\langle \pt \rangle$.
The systematic uncertainty related to the low \pt extrapolation is small
compared to the contributions from other sources and therefore is not included
in the combined systematic uncertainty of the measurement.

The tracker acceptance and the track reconstruction efficiency generally have
small uncertainties (1 and 3\%, respectively), but at very low \pt they reach
6\%.
For the multiple-track and misreconstructed-track rate corrections, the
uncertainty is assumed to be 50\% of the correction, while for the correction
for secondary particles it is estimated to be 25\% based on the differences
between predictions of MC event generators and data.
These bin-by-bin, largely uncorrelated uncertainties are caused by the
imperfect modeling of the detector: regions with incorrectly modeled tracking
efficiency, alignment uncertainties, and channel-by-channel varying hit
efficiency.
All these effects are taken as uncorrelated.

The statistical uncertainties in the extracted yields are given by the fit
uncertainties.
Variations of the track-level correction parameters, incompatible with
statistical fluctuations, are observed. They are used to estimate the
systematic uncertainties in the fitted scale factors and shifts and are at the
level of $10^{-2}$ and $2 \times10^{-3}$, respectively.
The systematic uncertainties in the yields in each bin are thus obtained by
refitting the histograms with the parameters changed by these amounts.
For the present measurement, systematic uncertainties dominate over the
statistical ones.

The systematic uncertainties originating from the unfolding procedure are also
studied. Since the \pt response matrices are close to diagonal, the unfolding
of the \pt distributions does not introduce substantial uncertainties.
The correlations between neighboring \pt bins are neglected, and therefore
statistical uncertainties are regarded as uncorrelated.
The systematic uncertainty of the fitted yields is in the range 1\%--10\%,
depending primarily on total momentum.

\begin{table*}
 \topcaption{Summary of the systematic uncertainties affecting the \pt
spectra. Values in parentheses indicate uncertainties in the
$\langle\pt\rangle$ measurement.
Representative, particle-specific uncertainties (\Pgp, \PK, \Pp) are given for
$\pt =0.6\GeVc$ in the third group of systematic uncertainties.}
 \label{tab:error}
 \centering
 \begin{scotch}{lcccc}
  \multirow{2}{*}{Source} & Uncertainty & \multicolumn{3}{c}{Propagated} \\
  & {of the source} [\%] & \multicolumn{3}{c}{yield uncertainty} [\%] \\
  \hline
  \multicolumn{3}{l}{Fully correlated, normalization} \\
  \; Correction for event selection & 3.0 (1.0)	& \multirow{3}{*}{$\biggl\}$} &
\multirow{3}{*}{3--4 (5--9)} & \multirow{3}{*}{} \\
  \; Pileup correction (merged and split vertices) & 0.3 & & & \\
  \; High-\pt extrapolation  & 1--3 (4--8) & & & \\
  \hline
  \multicolumn{3}{l}{Mostly uncorrelated} \\
  \; Pixel hit efficiency		& 0.3 & \multirow{2}{*}{$\biggl\}$} &
\multirow{2}{*}{0.3} & \\
  \; Misalignment, different scenarios	& 0.1 & & & \\
  \hline
  \multicolumn{2}{l}{Mostly uncorrelated, $(y,\pt)$-dependent} & \Pgp & \PK & \Pp \\
  \; Acceptance of the tracker	        & 1--6 & 1 & 1 & 1 \\
  \; Efficiency of the reconstruction	& 3--6 & 3 & 3 & 3 \\
  \; Multiple-track reconstruction      & 50\% of the corr.
                                         & \NA & \NA & \NA \\
  \; Misreconstructed-track rate	 & 50\% of the corr.
                                         & 0.1 & 0.1 & 0.1\\
  \; Correction for secondary particles  & 25\% of the corr.
                                         & 0.2  & \NA & 2 \\
  \; Fit of the $\mpe$ distributions & 1--10 & 1    & 2    & 1  \\
 \end{scotch}
\end{table*}

\section{Results}

\label{sec:results}

\begin{figure}
\centering
  \includegraphics[width=0.49\textwidth]{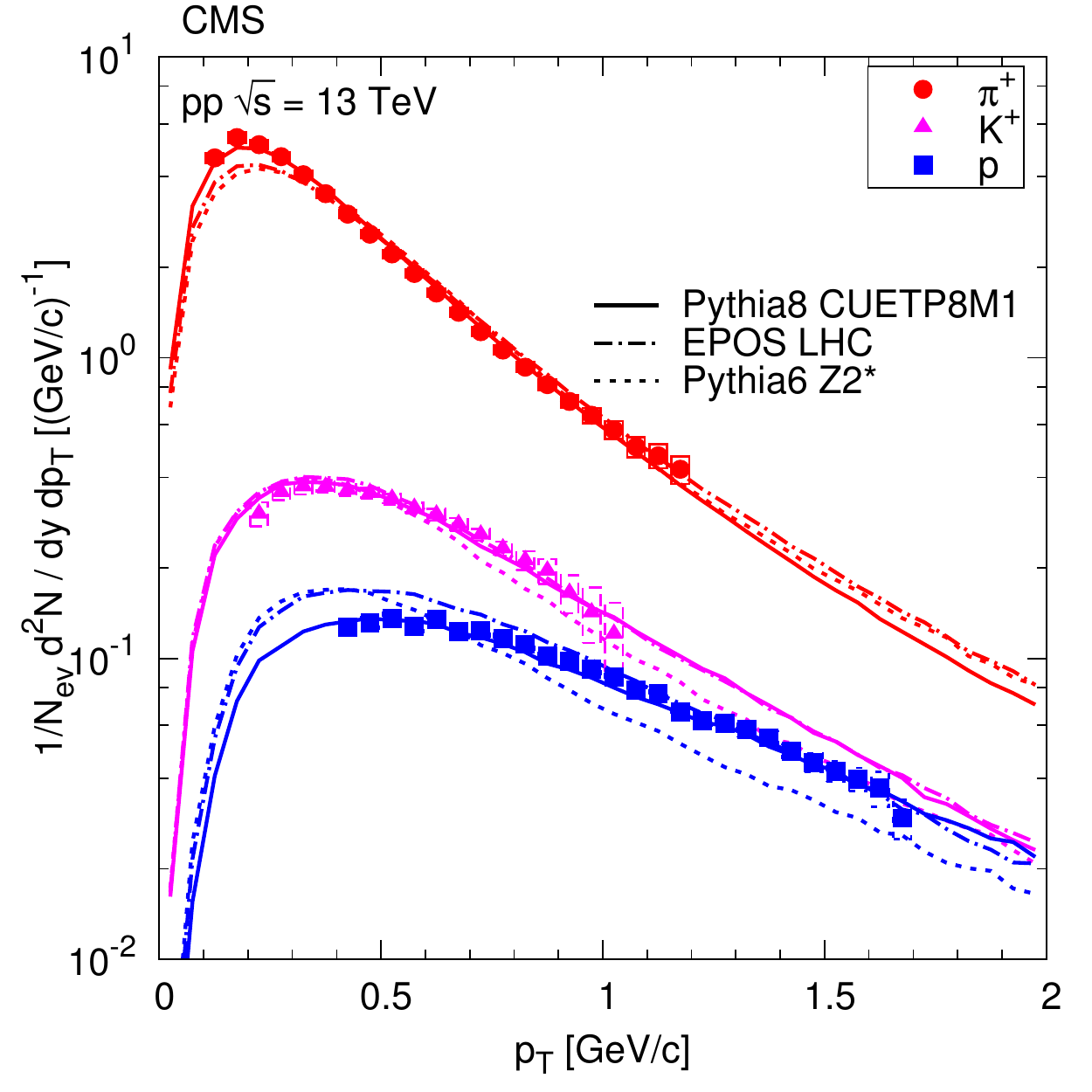}
  \includegraphics[width=0.49\textwidth]{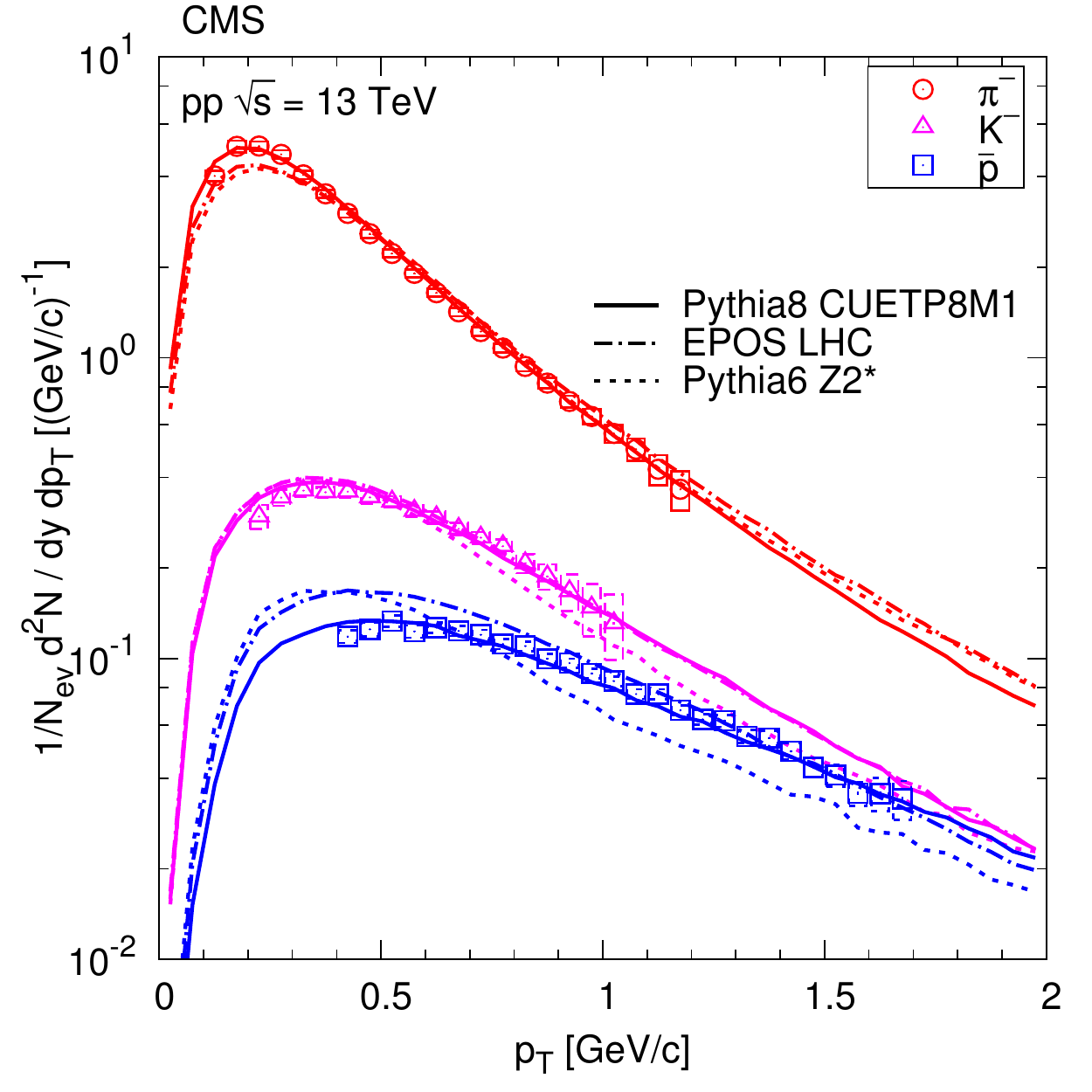}
 \caption{Transverse momentum distributions of identified charged hadrons
(pions, kaons, protons) from inelastic pp collisions, in the range $\abs{y}<1$,
for positively (\cmsLeft) and negatively (\cmsRight) charged particles.
Measured values (same as in Fig.~\ref{fig:dndpt_lin}) are plotted together with
predictions from {\PYTHIA}8, \EPOS, and {\PYTHIA}6.
Boxes show the uncorrelated systematic uncertainties, while error bars indicate
the uncorrelated statistical uncertainties (hardly visible).
The fully correlated normalization uncertainty (not shown) is 3.0\%.}
 \label{fig:dndpt_log}
\end{figure}

The results discussed in the following are averaged over the rapidity range $\abs{y} <
1$. In all cases, error bars in the figures indicate the uncorrelated
statistical uncertainties, while boxes show the uncorrelated systematic
uncertainties. The fully correlated normalization uncertainty is not shown. For
the \pt spectra, the average transverse momentum $\langle\pt\rangle$, and the
ratios of particle yields, the data are compared to the predictions of
{\PYTHIA}8, \EPOS, and {\PYTHIA}6.

\subsection{Inclusive measurements}

The transverse momentum distributions of positively and negatively charged
hadrons (pions, kaons, protons) are shown in Fig.~\ref{fig:dndpt_lin}, along
with the results of the fits to the Tsallis-Pareto parametrization
[Eqs.~\eqref{eq:tsallis} and \eqref{eq:tsallis2}]. The fits are of good quality
with $\chi^2/\mathrm{ndf}$ values in the range
0.4--1.2 (Table~\ref{tab:fits}). Figure~\ref{fig:dndpt_log} presents the same
data compared to the {\PYTHIA}8, \EPOS, and {\PYTHIA}6 predictions. While pions
are described well by all three generators, kaons are best modelled by
{\PYTHIA}8 and \EPOS. For protons and very low \pt pions only {\PYTHIA}8
gives a good description of the data.

\begin{table*}
 \topcaption{Fit results for $\rd N/\rd y$, $n$, and $T$ (obtained via
Eqs.~\eqref{eq:tsallis} and \eqref{eq:tsallis2}), associated goodness-of-fit
values, and extracted $\langle \rd N/\rd y \rangle$ and $\langle\pt\rangle$
averages, for charged pion, kaon, and proton spectra measured in the range
$\abs{y} < 1$ in inelastic pp collisions at 13\TeV.
Combined statistical and systematic uncertainties are given.}
 \label{tab:fits}
 \centering
 \begin{scotch}{ccrccccc}
  {Particle} & $\rd N/\rd y$ & \multicolumn{1}{c}{$n$} & $T$ [$\GeVcns{}$]
             & $\chi^2/\mathrm{ndf}$
             & $\langle \rd N/\rd y \rangle$
             & $\langle \pt \rangle$ [$\GeVcns{}$] \\
 \hline
\Pgpp & 2.833 $\pm$ 0.031 & 5.2 $\pm$ 0.2 & 0.119 $\pm$ 0.003 & 6.8/19 & 2.843 $\pm$ 0.034 & 0.51 $\pm$ 0.03 \\
\Pgpm & 2.733 $\pm$ 0.029 & 5.9 $\pm$ 0.2 & 0.130 $\pm$ 0.003 &  22/19 & 2.746 $\pm$ 0.031 & 0.50 $\pm$ 0.03 \\
\PKp  & 0.318 $\pm$ 0.021 & 15  $\pm$ 18  & 0.231 $\pm$ 0.025 & 7.3/14 & 0.318 $\pm$ 0.007 & 0.67 $\pm$ 0.03 \\
\PKm  & 0.332 $\pm$ 0.026 & 7.7 $\pm$ 4.6 & 0.217 $\pm$ 0.024 & 5.0/14 & 0.331 $\pm$ 0.011 & 0.75 $\pm$ 0.05 \\
\Pp   & 0.169 $\pm$ 0.007 & 4.7 $\pm$ 0.8 & 0.222 $\pm$ 0.016 & 8.9/23 & 0.169 $\pm$ 0.004 & 1.10 $\pm$ 0.12 \\
\Pap  & 0.162 $\pm$ 0.006 & 5.3 $\pm$ 1.1 & 0.237 $\pm$ 0.016 & 8.4/23 & 0.162 $\pm$ 0.004 & 1.07 $\pm$ 0.09 \\
 \end{scotch}
\end{table*}

\begin{figure}
 \centering
  \includegraphics[width=0.49\textwidth]{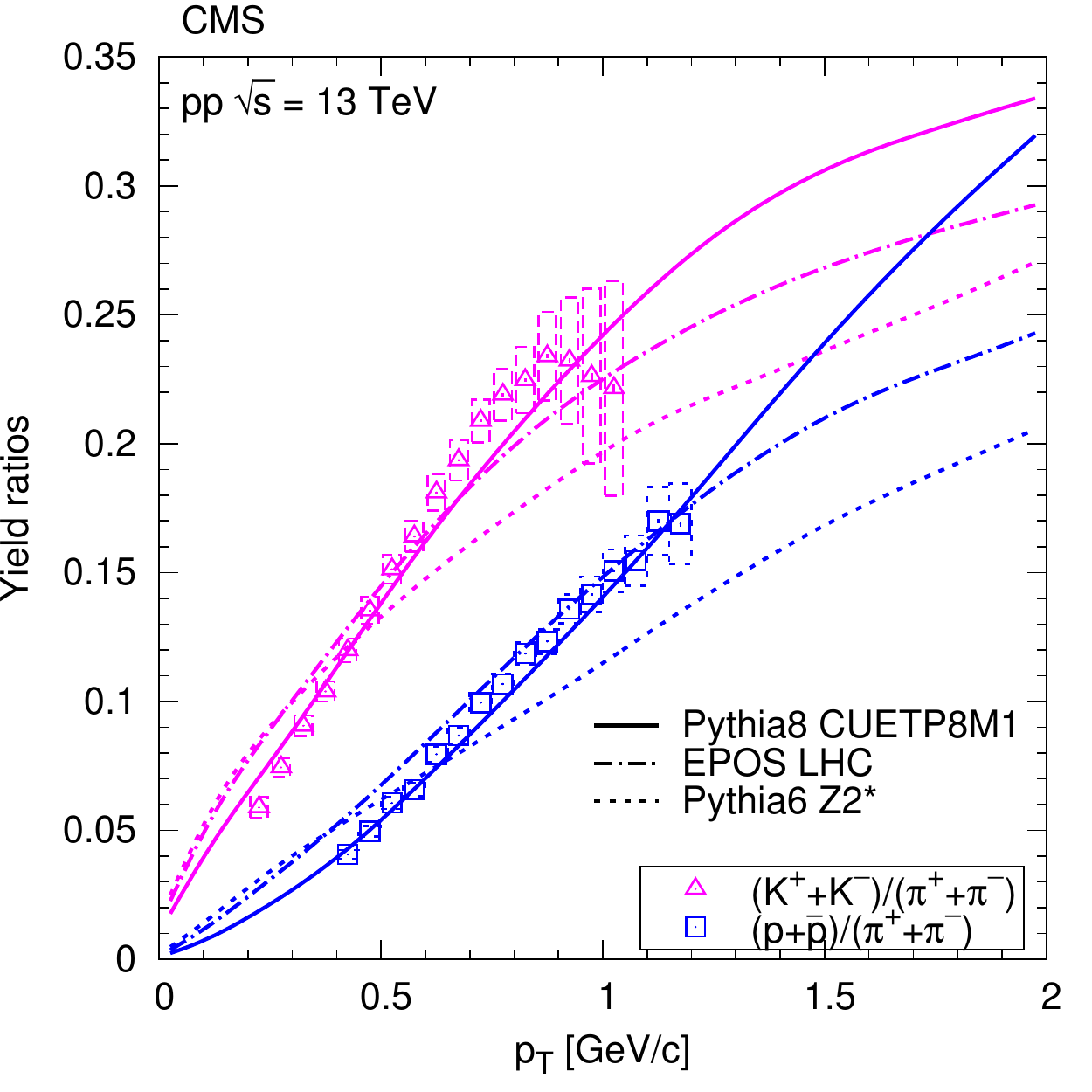}
  \includegraphics[width=0.49\textwidth]{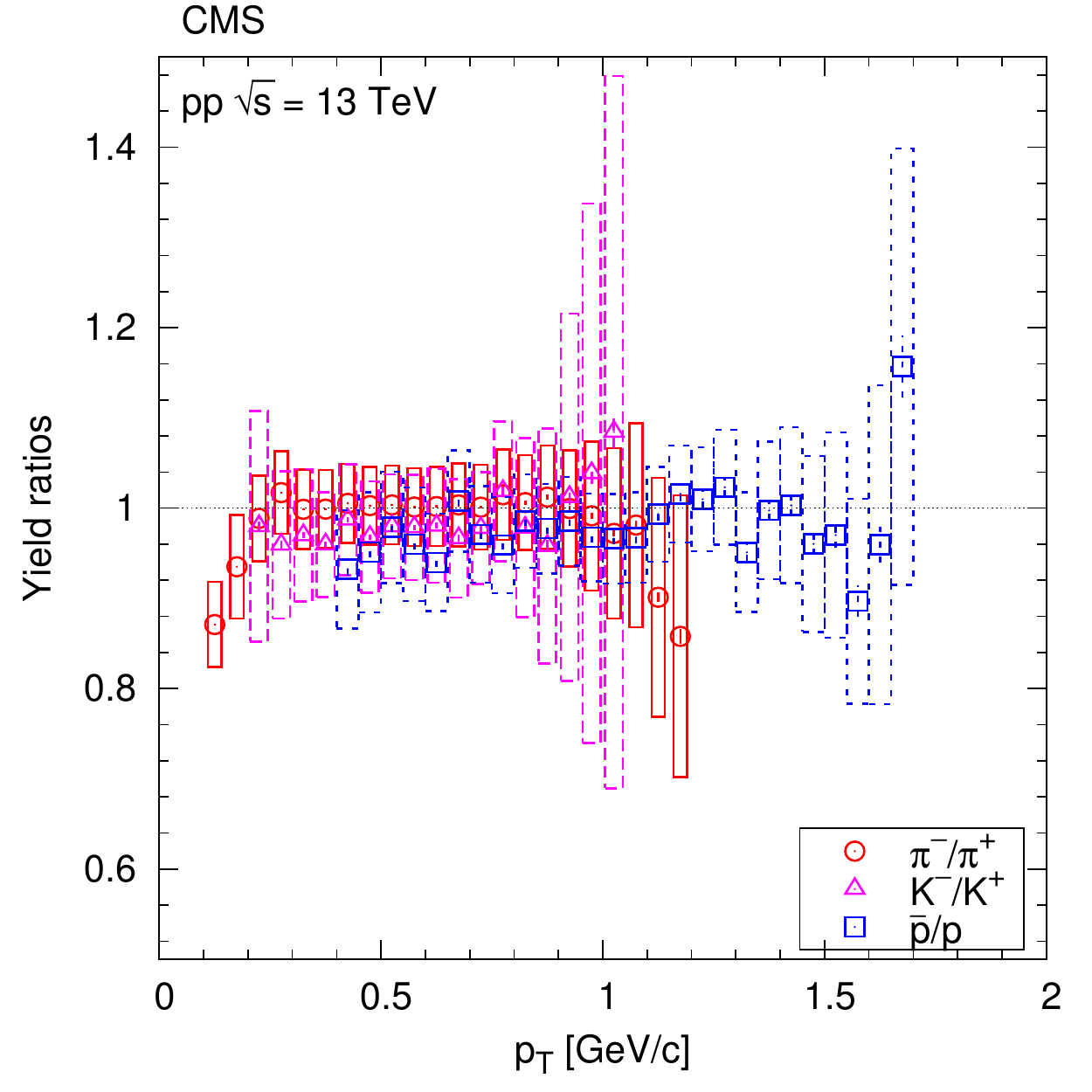}
 \caption{Ratios of particle yields, \PK/\Pgp\ and \Pp/\Pgp\ (left) and
opposite-charge ratios (right), as a function of transverse momentum.
Error bars indicate the uncorrelated statistical uncertainties, while boxes
show the uncorrelated systematic uncertainties. In the \cmsLeft panel, curves
indicate predictions from {\PYTHIA}8, \EPOS, and {\PYTHIA}6.}
 \label{fig:ratios_vs_pt}
\end{figure}

Ratios of particle yields as a function of the transverse momentum are plotted
in Fig.~\ref{fig:ratios_vs_pt}.
Only {\PYTHIA}8 is able to predict both the $\PK/\Pgp$ and $\Pp/\Pgp$ ratios as
a function of \pt.
The ratios of the yields for oppositely charged particles are close to one
(Fig.~\ref{fig:ratios_vs_pt}, right), as
expected at this center-of-mass energy in the central rapidity region.

\subsection{Multiplicity-dependent measurements}

The study of the \pt spectra as a function of the event track multiplicity is
motivated partly by the intriguing hadron correlations measured in pp and pPb
collisions at high track
multiplicities~\cite{Khachatryan:2010gv,CMS:2012qk,Abelev:2012ola,Aad:2012gla},
suggesting possible collective effects in ``central'' collisions at the LHC\@.
We have also observed that in pp collisions at LHC
energies~\cite{identifiedSpectra,Chatrchyan:2013eya}, the characteristics of
particle production ($\langle \pt \rangle$, ratios of yields) are strongly
correlated with the particle multiplicity in the event, which is in itself
closely related to the number of underlying parton-parton interactions,
independently of the concrete center-of-mass energy of the pp collision.

\begin{figure*}
 \centering
  \includegraphics[width=0.49\textwidth]{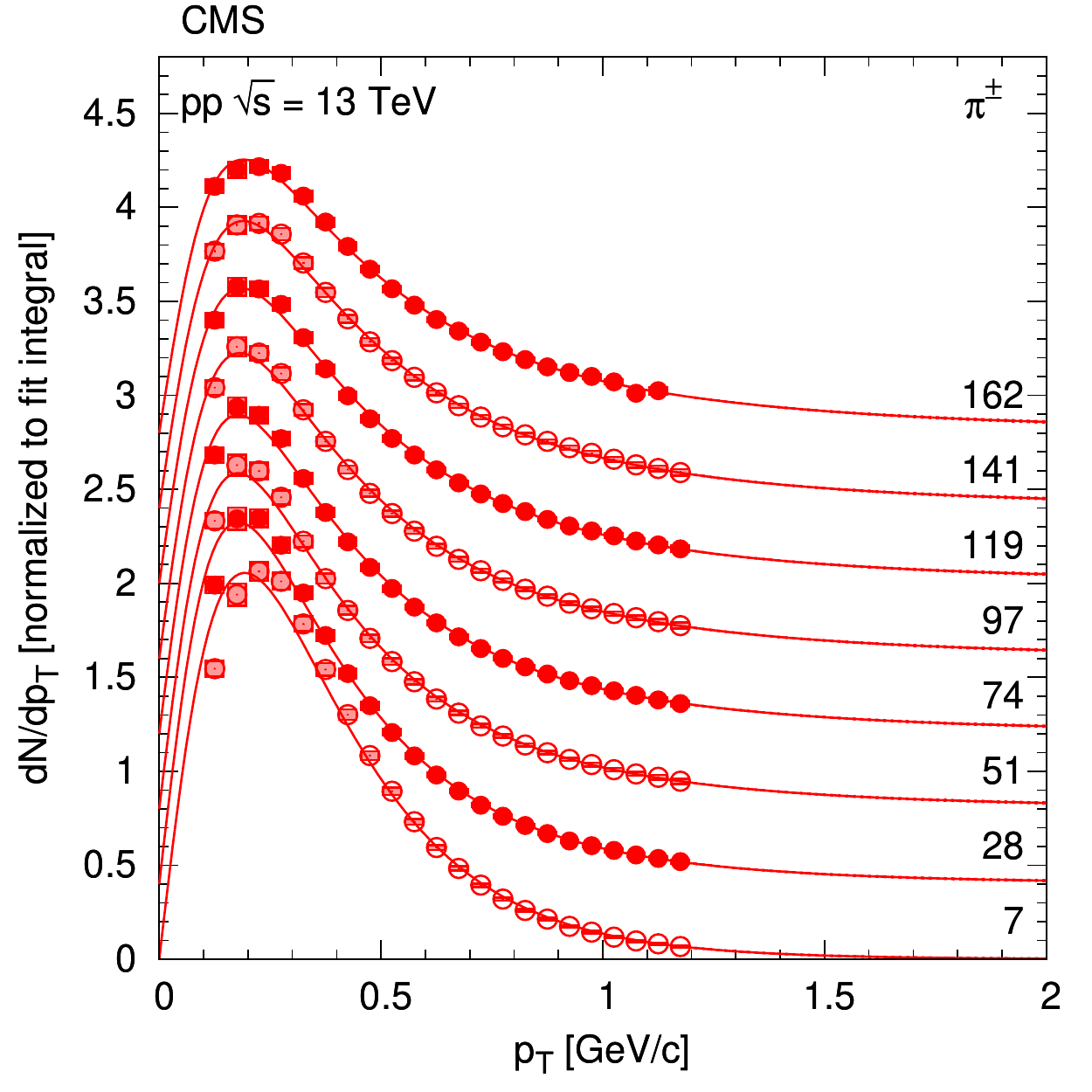}
  \includegraphics[width=0.49\textwidth]{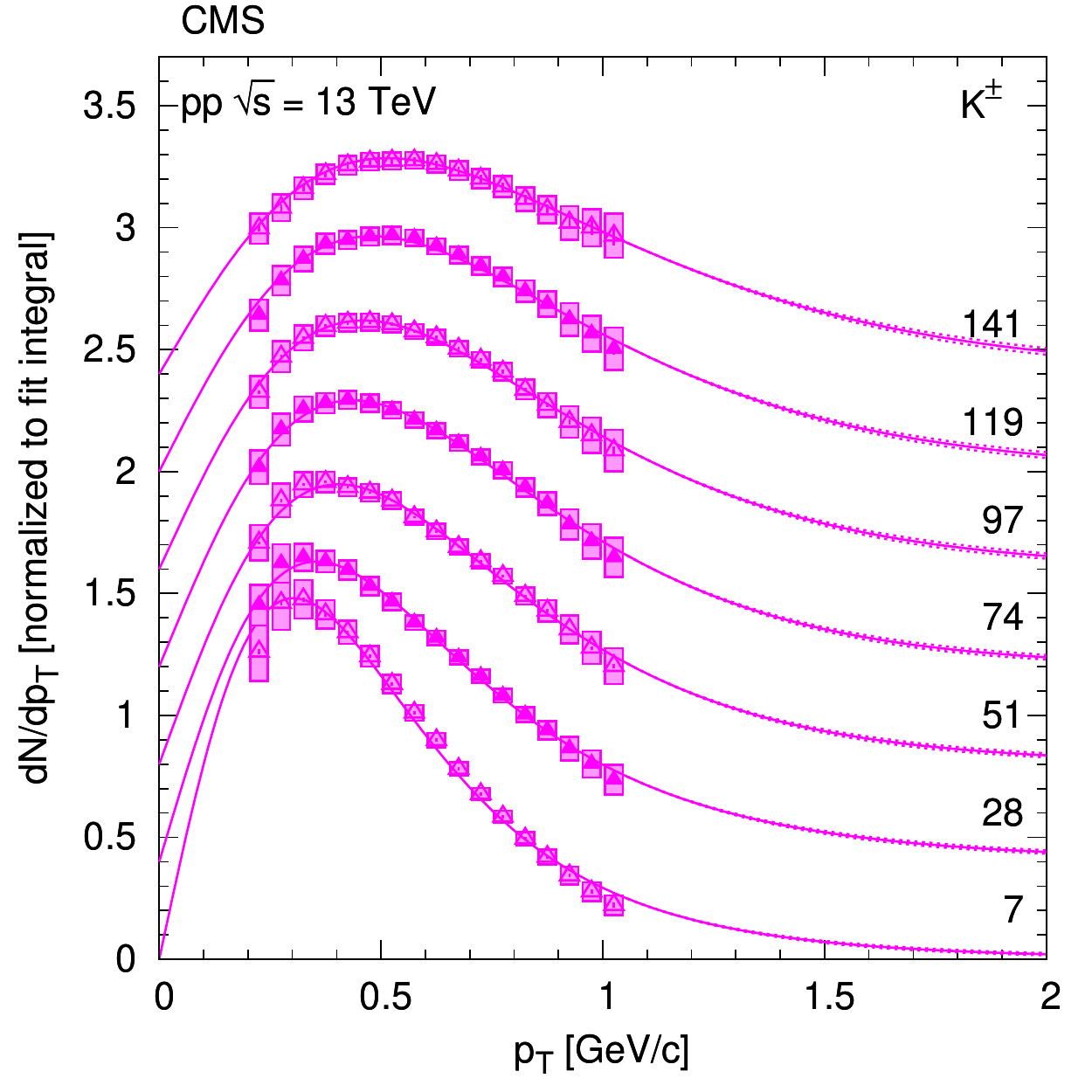}
  \includegraphics[width=0.49\textwidth]{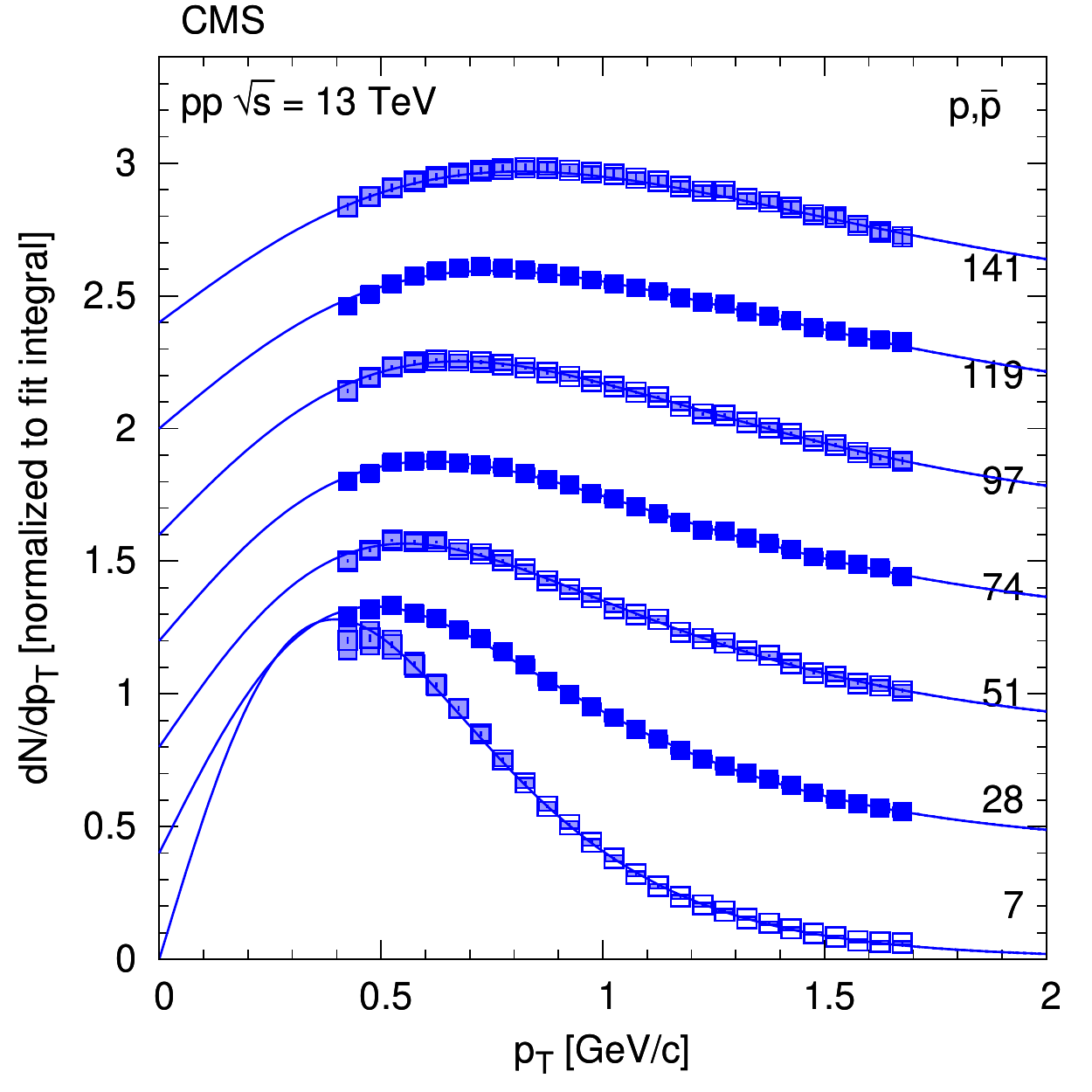}
 \caption{Transverse momentum distributions of charged pions (top left), kaons
(top right), and protons (bottom), normalized such that the fit integral is
unity, in every selected multiplicity class ($\langle N_\text{tracks} \rangle$
values are indicated) in the range $\abs{y}<1$, fitted with the Tsallis-Pareto
parametrization (solid lines).
For better visibility, the result for any given $\langle N_\text{tracks}
\rangle$ bin is shifted by 0.4 units with respect to the adjacent bins.
Error bars indicate the uncorrelated statistical uncertainties, while boxes
show the uncorrelated systematic uncertainties.
Dotted lines (mostly indistinguishable from the nominal fit curves) illustrate
the effect of varying the inverse exponent ($1/n$) of the Tsallis-Pareto
function by $\pm$0.05 beyond the highest-\pt measured point.}
 \label{fig:dndpt_multi}
\end{figure*}

\begin{table*}
 \topcaption{Relationship between the number of reconstructed tracks
($N_\text{rec}$) and the average number of corrected tracks ($\langle
N_\text{tracks} \rangle$) in the region $\abs{\eta} < 2.4$ in the 18
multiplicity classes considered.}
 \label{tab:multiClass}
 \centering
 \begin{scotch}{cc@{\hspace{\sptw}}c@{\hspace{\sptw}}c@{\hspace{\sptw}}c@{\hspace{\sptw}}c@{\hspace{\sptw}}c@{\hspace{\sptw}}c@{\hspace{\sptw}}c@{\hspace{\sptw}}c@{\hspace{\sptw}}c@{\hspace{\sptw}}c@{\hspace{\sptw}}c@{\hspace{\sptw}}c@{\hspace{\sptw}}c@{\hspace{\sptw}}c@{\hspace{\sptw}}c@{\hspace{\sptw}}c@{\hspace{\sptw}}c}
\\[-1.5ex]
  $N_\text{rec}$ &
  \begin{sideways}   0--9   \end{sideways} &
  \begin{sideways}  10--19  \end{sideways} &
  \begin{sideways}  20--29  \end{sideways} &
  \begin{sideways}  30--39  \end{sideways} &
  \begin{sideways}  40--49  \end{sideways} &
  \begin{sideways}  50--59  \end{sideways} &
  \begin{sideways}  60--69  \end{sideways} &
  \begin{sideways}  70--79  \end{sideways} &
  \begin{sideways}  80--89  \end{sideways} &
  \begin{sideways}  90--99  \end{sideways} &
  \begin{sideways} 100--109 \end{sideways} &
  \begin{sideways} 110--119 \end{sideways} &
  \begin{sideways} 120--129 \end{sideways} &
  \begin{sideways} 130--139 \end{sideways} &
  \begin{sideways} 140--149 \end{sideways} &
  \begin{sideways} 150--159 \end{sideways} &
  \begin{sideways} 160--169 \end{sideways} &
  \begin{sideways} 170--179 \end{sideways} \\
  \hline
  $\langle N_\text{tracks} \rangle$ &
     7 &  16 &  28 &  40 &  51 &  63 &  74 &  85 &  97 & 108 &
   119 & 130 & 141 & 151 & 162 & 172 & 183 & 187 \\
 \end{scotch}
\end{table*}

\begin{figure}
 \centering
  \includegraphics[width=0.49\textwidth]{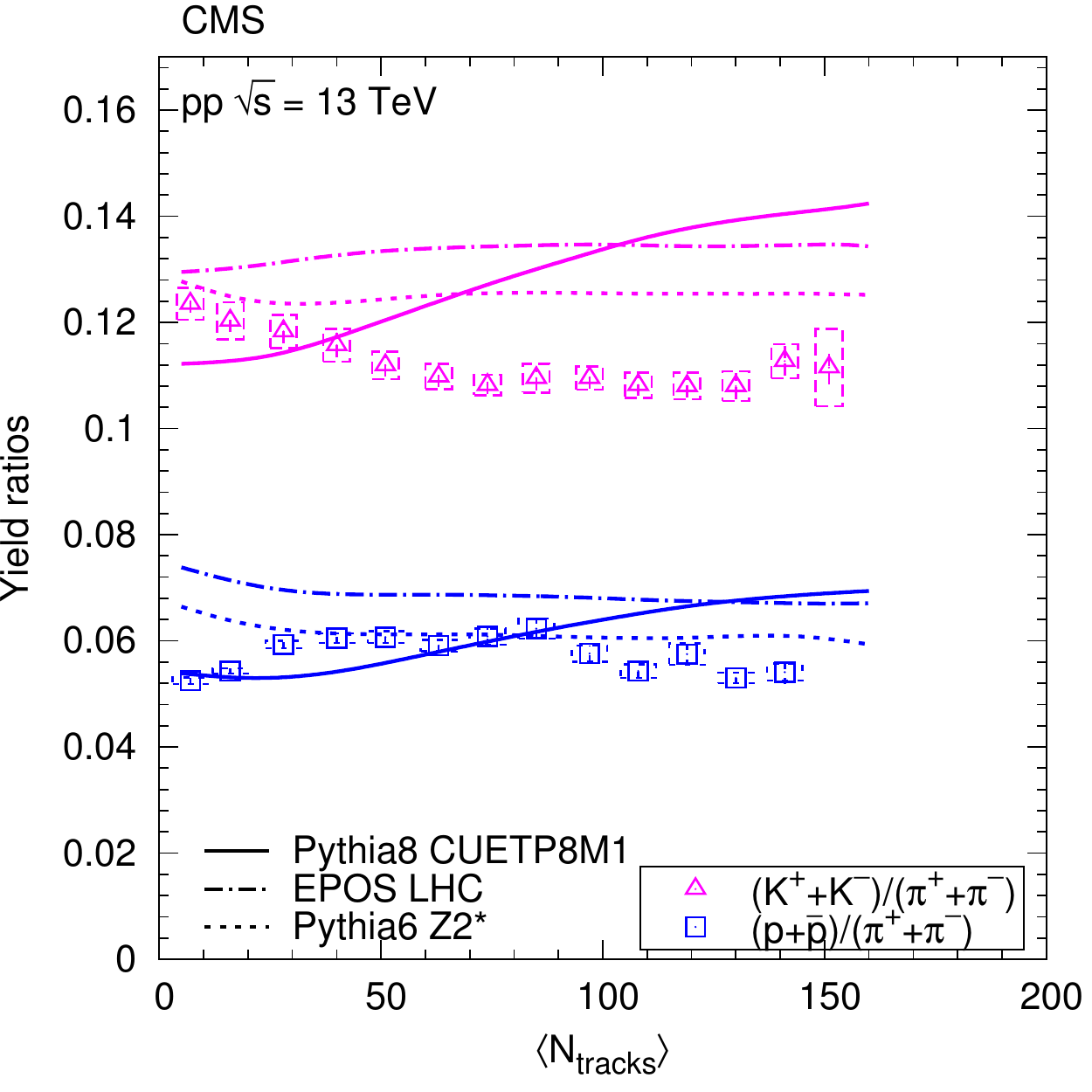}
  \includegraphics[width=0.49\textwidth]{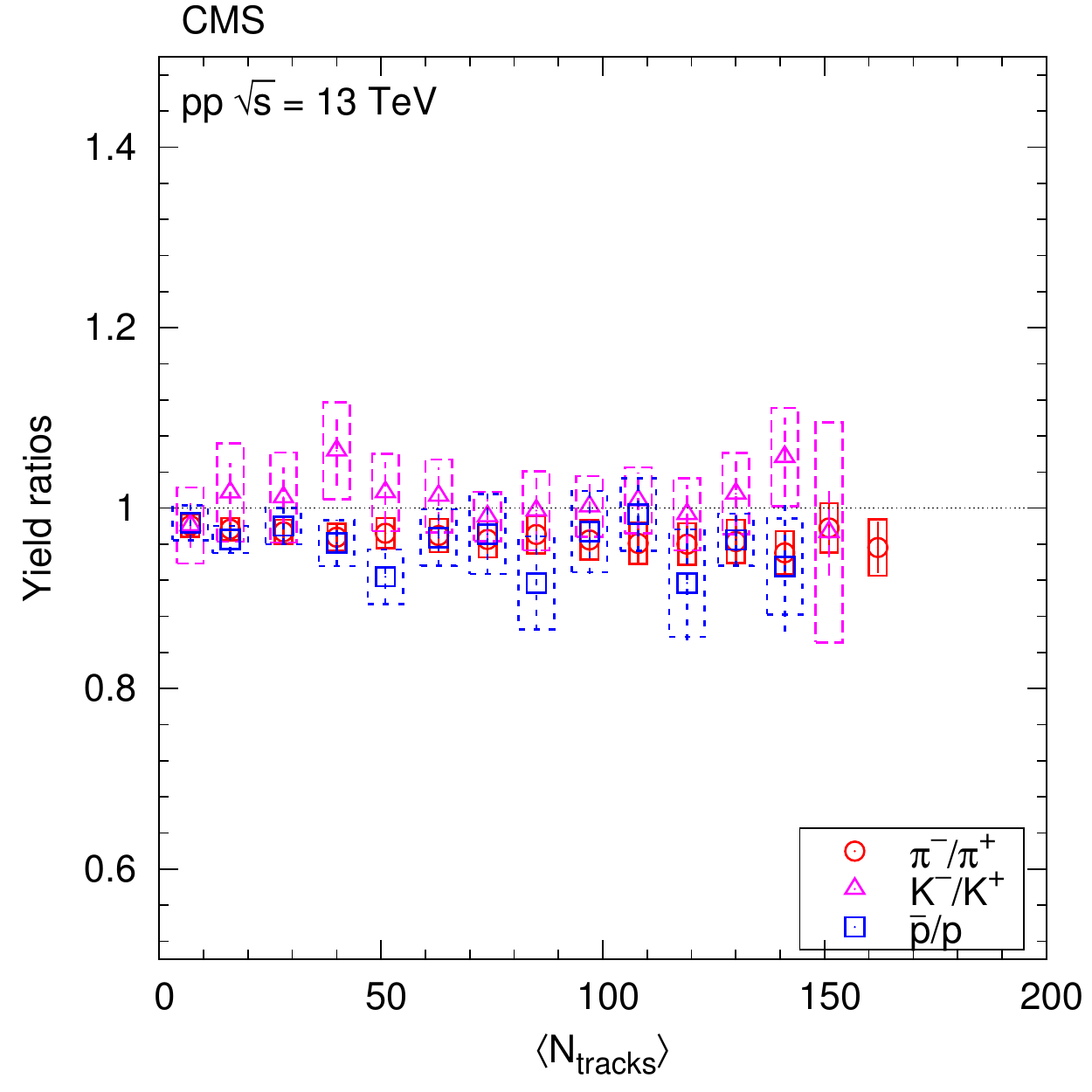}
 \caption{Ratios of particle yields in the range $\abs{y}<1$ as a function of
the corrected track multiplicity for $\abs{\eta}<2.4$.
The \PK/\Pgp\ and \Pp/\Pgp\ values are shown in the \cmsLeft panel, and
opposite-charge ratios are plotted in the \cmsRight panel.
Error bars indicate the uncorrelated combined uncertainties, while boxes show
the uncorrelated systematic uncertainties. In the \cmsLeft panel, curves
indicate predictions from {\PYTHIA}8, \EPOS, and {\PYTHIA}6.}
 \label{fig:ratios_vs_multi}
\end{figure}

\begin{figure}
 \centering
  \includegraphics[width=0.49\textwidth]{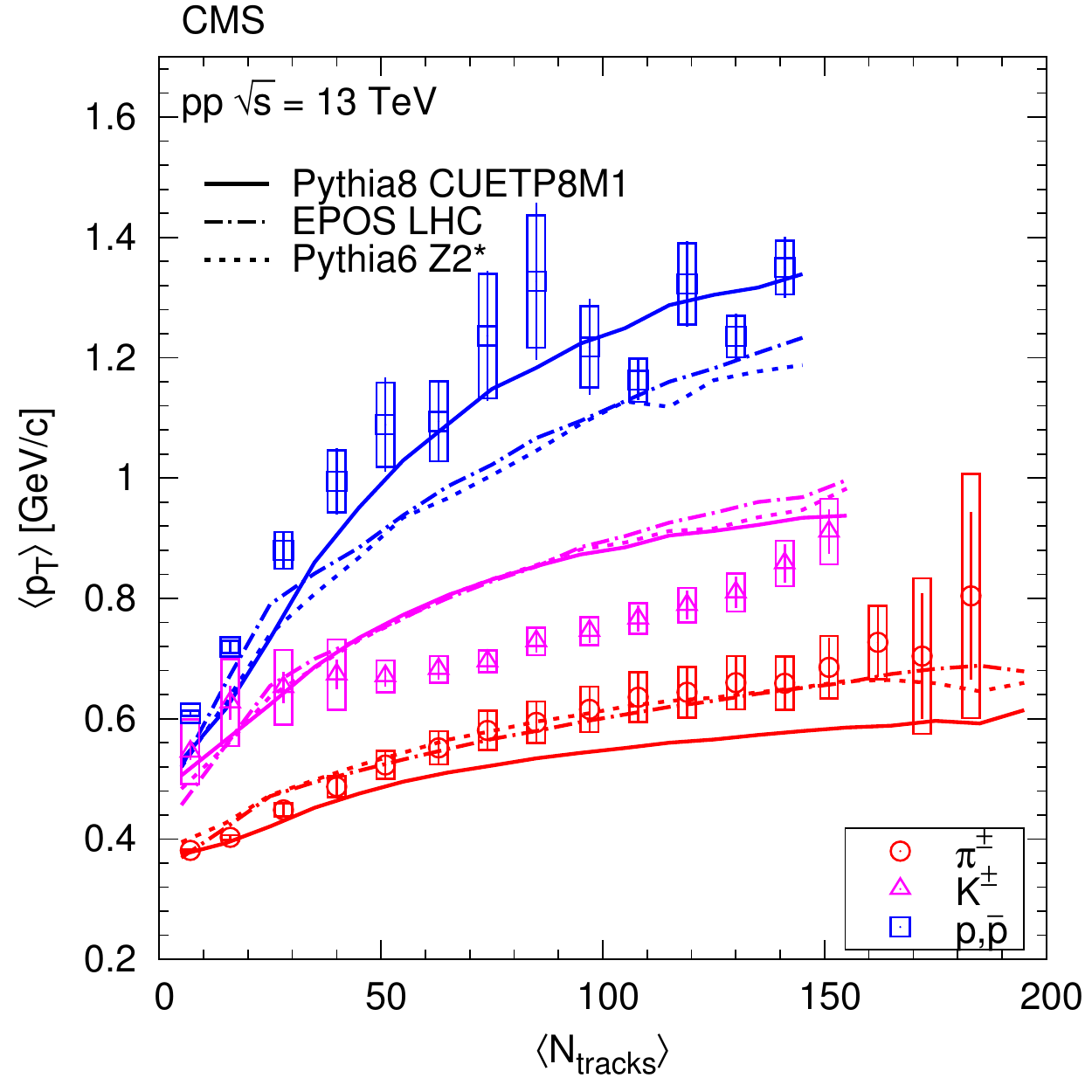}
 \caption{Average transverse momentum of identified charged hadrons (pions,
kaons, protons) in the range $\abs{y}<1$, as functions of the corrected track
multiplicity for $\abs{\eta}<2.4$, computed assuming a Tsallis-Pareto
distribution in the unmeasured range. Error bars indicate the uncorrelated
combined uncertainties, while boxes show the uncorrelated systematic
uncertainties. The fully correlated normalization uncertainty (not shown) is
1.0\%. Curves indicate predictions from {\PYTHIA}8, \EPOS, and {\PYTHIA}6.}
 \label{fig:apt_vs_multi}
\end{figure}

\begin{figure}
 \centering
  \includegraphics[width=0.49\textwidth]{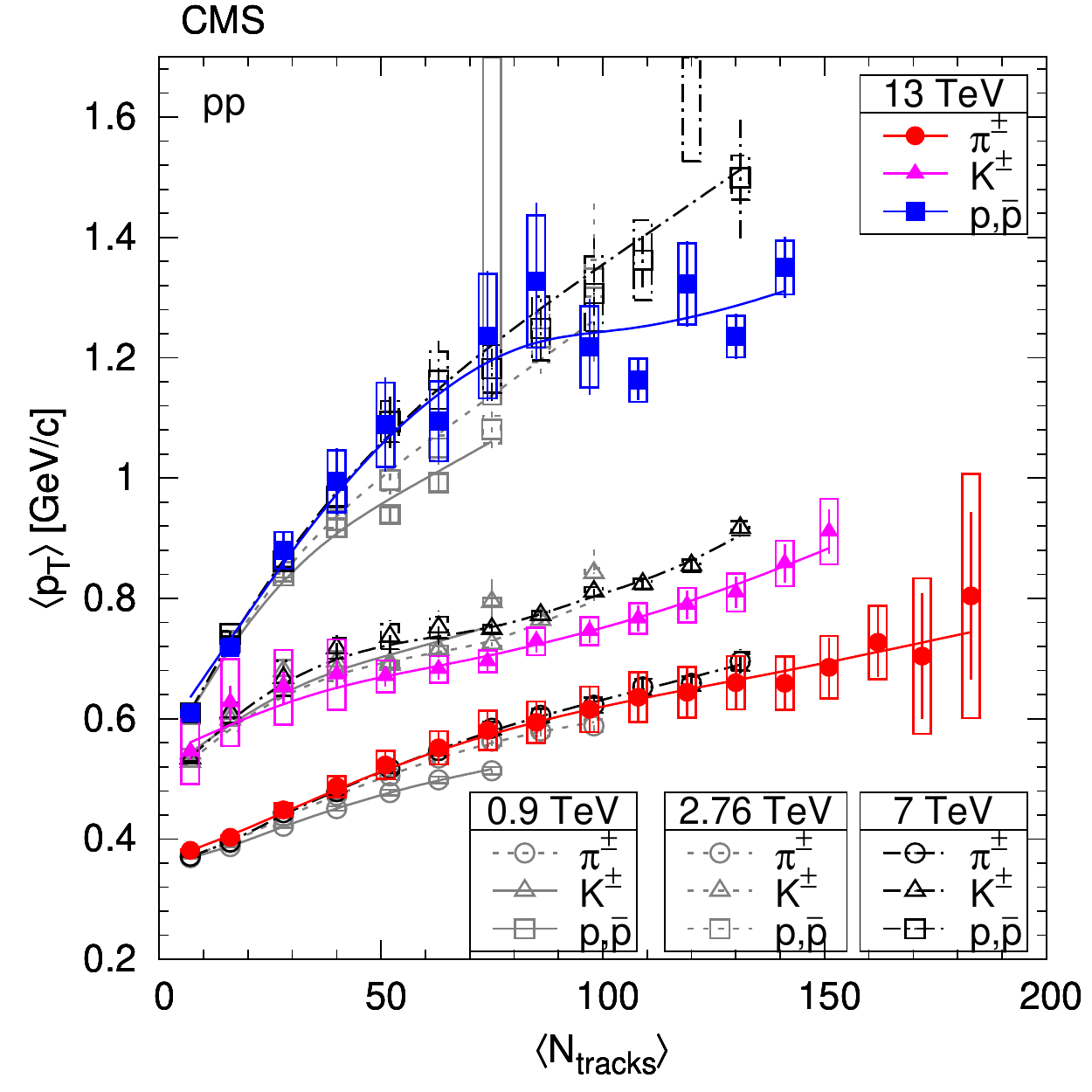}
  \includegraphics[width=0.49\textwidth]{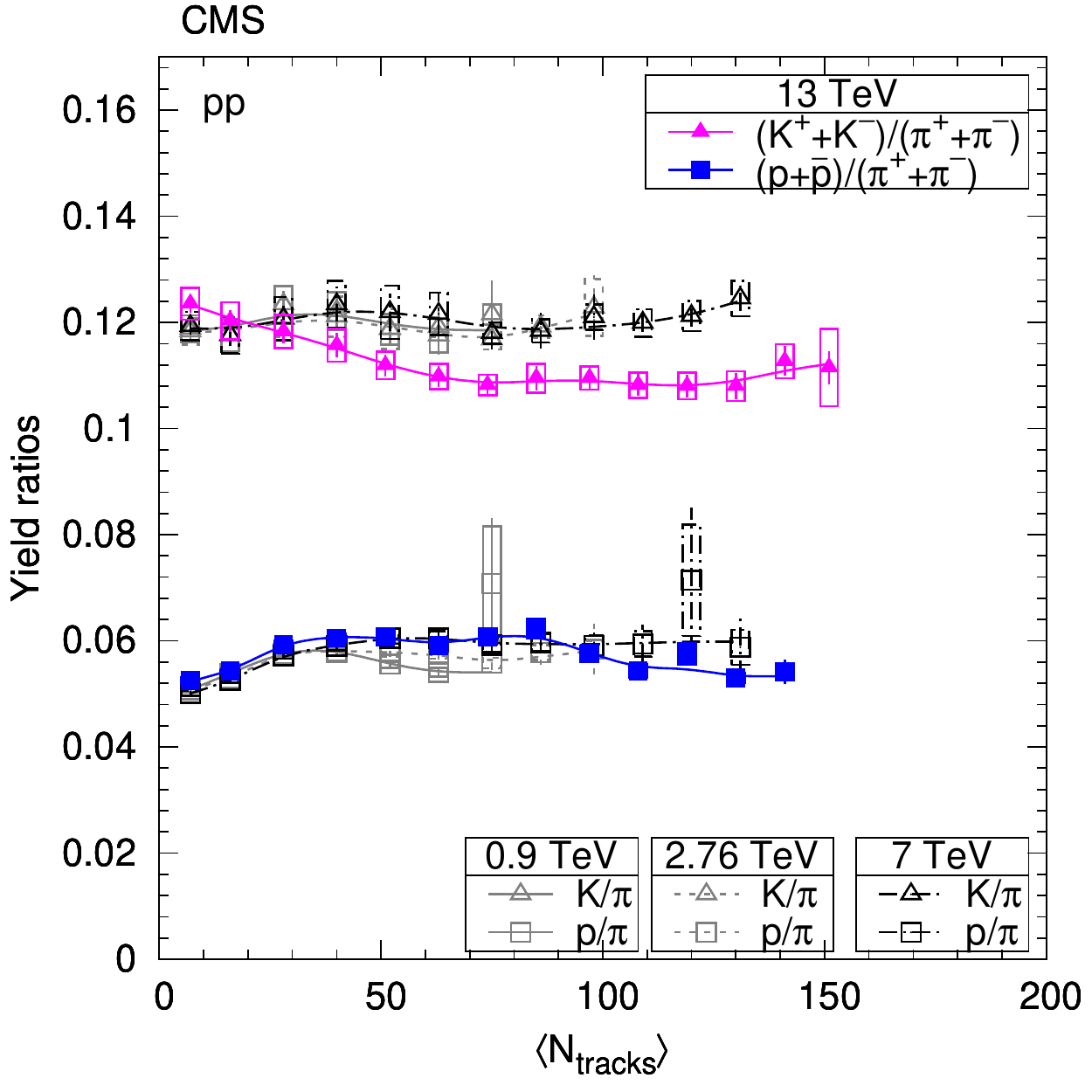}
 \caption{Average transverse momentum of identified charged hadrons (pions,
kaons, protons; \cmsLeft panel) and ratios of particle yields (\cmsRight panel)
in the range $\abs{y}<1$ as functions of the corrected track multiplicity for
$\abs{\eta}<2.4$, for pp collisions at $\sqrt{s} =13$\TeV (filled symbols) and
at lower energies (open symbols)~\cite{identifiedSpectra}.
Both $\langle\pt\rangle$ and yield ratios are computed assuming a
Tsallis-Pareto distribution in the unmeasured range.
Error bars indicate the uncorrelated combined uncertainties, while boxes show
the uncorrelated systematic uncertainties. For $\langle\pt\rangle$ the fully
correlated normalization uncertainty (not shown) is 1.0\%.
In both plots, lines are drawn to guide the eye (gray solid: 0.9\TeV, gray
dotted: 2.76\TeV, black dash-dotted: 7\TeV, colored solid: 13\TeV).}
 \label{fig:multiplicityDependence}
\end{figure}

The event track multiplicity, $N_\text{rec}$, is defined as the number of
tracks with $\abs{\eta} < 2.4$ reconstructed using the same algorithm as for the
identified charged hadrons~\cite{Sikler:2007uh}.
The event multiplicity is divided into 18 classes as defined in
Table~\ref{tab:multiClass}.
To facilitate comparisons with models, the event charged-particle multiplicity
over $\abs{\eta} < 2.4$ ($N_\text{tracks}$) is determined for each multiplicity
class by correcting $N_\text{rec}$ for the track reconstruction efficiency,
which is estimated with the {\PYTHIA}8 simulation in $(\eta,\pt)$ bins.
The corrected yields are then integrated over \pt, down to zero yield at $\pt =
0$ (with a linear extrapolation below $\pt = 0.1\GeVc$). Finally, the
integrals for each $\eta$ slice are summed up.
The average corrected charged-particle multiplicity $\langle N_\text{tracks}
\rangle$ is shown in Table~\ref{tab:multiClass} for each event multiplicity
class. The value of $\langle N_\text{tracks} \rangle$ is used to identify the
multiplicity class in
Figs.~\ref{fig:dndpt_multi}--\ref{fig:multiplicityDependence}.

Transverse-momentum distributions of pions, kaons, and protons, measured over
$\abs{y} < 1$ and normalized such that the fit integral is unity, are shown in
Fig.~\ref{fig:dndpt_multi} for various multiplicity classes.
The distributions of negatively and positively charged particles are summed.
The Tsallis-Pareto parametrization is fitted to the distributions with
$\chi^2/\mathrm{ndf}$ values in the range
0.3--2.3 for pions, 0.2--2.6 for kaons, and 0.1--0.8 for protons.
It is observed that for kaons and protons, the parameter $T$ increases with
multiplicity, while for pions $T$ slightly increases and the exponent $n$
slightly decreases with multiplicity.

The ratios of particle yields are displayed as functions of track multiplicity
in Fig.~\ref{fig:ratios_vs_multi}. The $\PK/\Pgp$ and $\Pp/\Pgp$ ratios are
relatively flat as a function of $\langle N_\text{tracks} \rangle$, and none of
the models is able to accurately reproduce the track multiplicity dependence.
The ratios of yields of oppositely charged particles are independent of
$\langle N_\text{tracks} \rangle$ as shown in the \cmsRight panel of
Fig.~\ref{fig:ratios_vs_multi}.
The average transverse momentum $\langle\pt\rangle$ is shown as a function of
multiplicity in Fig.~\ref{fig:apt_vs_multi}. Although {\PYTHIA}8 gives a good
description of the (multiplicity integrated) inelastic \pt spectra
(Fig.~\ref{fig:dndpt_log}), none of the MC event generators reproduces well the
multiplicity dependence of $\langle \pt \rangle$ for all particle species. In
particular, all generators overestimate the measured values for kaons. Pions
are well described by {\PYTHIA}6 and \EPOS, while protons are best described by
{\PYTHIA}8.

In the lower multiplicity events, with fewer than 50 tracks, we observe a
reasonable agreement between the data and the MC generator predictions for the
different particle yields. However in higher multiplicity events, the measured
kaon (proton) yield is smaller (higher) than predicted by the models. This
indicates that the MC parameters that control the strangeness and baryon
production as a function of parton multiplicity, need additional fine-tuning.

\subsection{Comparisons with lower energy pp data}
\label{sec:comparisons}

\begin{figure}
 \centering
  \includegraphics[width=0.49\textwidth]{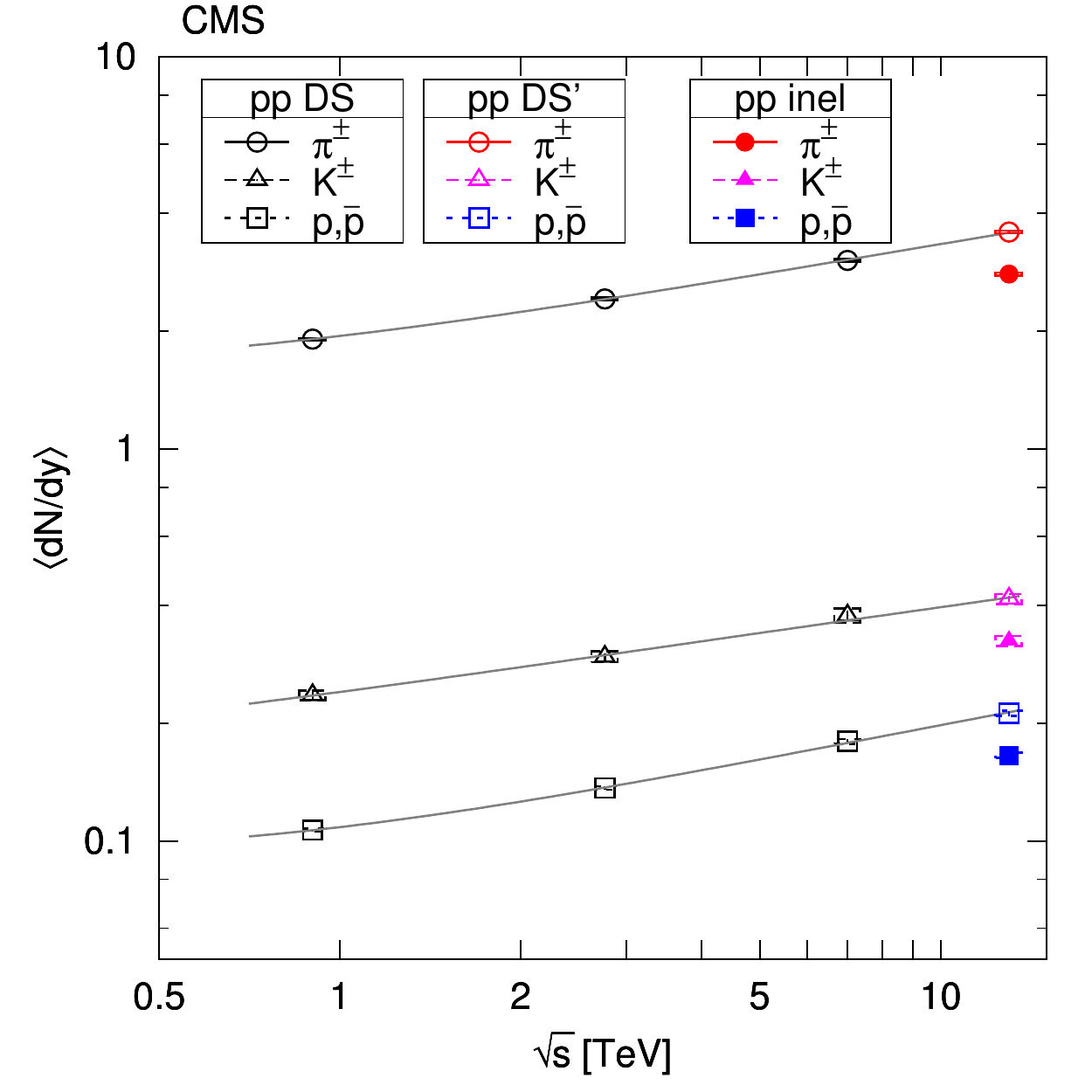}
  \includegraphics[width=0.49\textwidth]{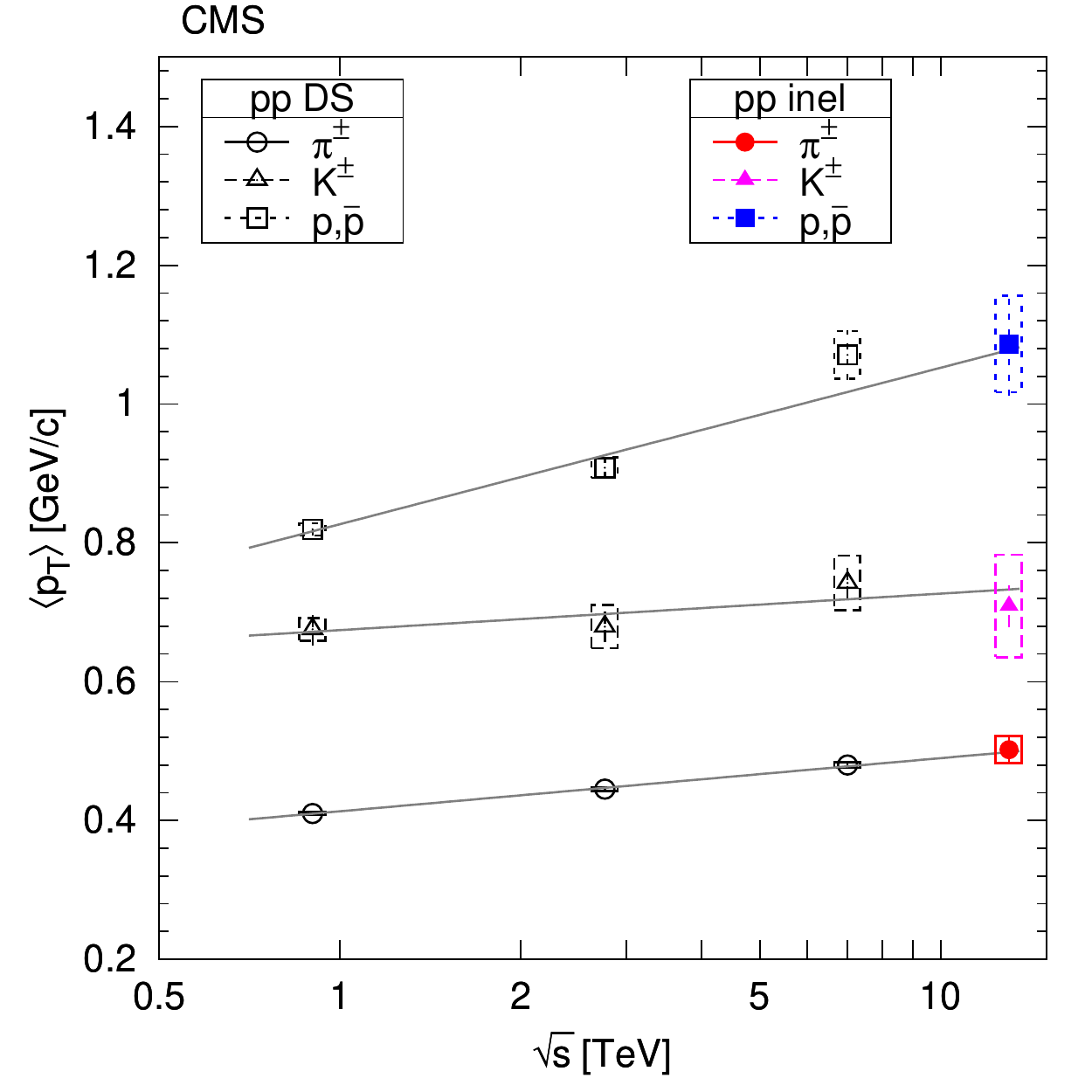}
 \caption{Average rapidity densities $\langle\rd N/\rd y\rangle$ (\cmsLeft) and
average transverse momenta $\langle \pt \rangle$ (\cmsRight) for $\abs{y}<1$ as
functions of center-of-mass energy for pp collisions (with data at 0.9, 2.76,
and 7\TeV~\cite{identifiedSpectra}), for charge-averaged pions, kaons, and
protons.
In the left plot the pp DS' results at 13\TeV have been extrapolated from the
inelastic values using simulation.
Error bars indicate the uncorrelated combined uncertainties, while boxes show
the uncorrelated systematic uncertainties. The curves show parabolic
($\langle\rd N/\rd y\rangle$) or linear (for $\langle \pt \rangle$) fits in
$\ln s$.}
 \label{fig:energyDep}
\end{figure}

The comparison of these results with lower-energy pp data taken at various
center-of-mass energies (0.9, 2.76, and 7\TeV)~\cite{identifiedSpectra} is
presented in Fig.~\ref{fig:multiplicityDependence}, where the
track-multiplicity dependence of $\langle\pt\rangle$ (left) and the particle
yield ratios ($\PK/\Pgp$ and $\Pp/\Pgp$, right) are shown.
In the previous publication~\cite{identifiedSpectra}, the final results are
corrected to a particle-level selection that requires at least one particle
(with proper lifetime $\tau > 10^{-18}\unit{s}$) with $E > 3\GeV$ in the range
$-5 < \eta < -3$ and at least one in the range $3 < \eta <5$. This selection is
referred to as the ``double-sided'' (DS) selection.
Average rapidity densities $\langle\rd N/\rd y\rangle$ and average transverse
momenta $\langle \pt \rangle$ of charge-averaged pions, kaons, and protons as a
function of center-of-mass energy are shown in Fig.~\ref{fig:energyDep}
corrected to the DS selection (pp DS'). Based on the predictions of the three
MC event generators studied, the inelastic $\langle\rd N/\rd y\rangle$ result
is corrected upwards by 28\%, with an additional systematic uncertainty of
about 2\%. No such correction is applied in the case of $\langle \pt \rangle$,
since the inelastic value is close to the DS' one, with a difference of about
1\%.

The average \pt increases with particle mass and event multiplicity at all
$\sqrt{s}$, as
predicted by all considered event generators.
We note that both $\langle\pt\rangle$ and ratios of hadron yields show very
similar dependences on the particle multiplicity in the event, independently of
the center-of-mass energy of the pp collisions. The $\sqrt{s}$ evolution of the
average hadron \pt provides useful information on the so-called ``saturation
scale'' ($Q_\text{sat}$) of the gluons in the proton~\cite{dEnterria:2016oxo}.
Minijet-based models such as {\PYTHIA} have an energy-dependent infrared \pt
cutoff of the perturbative multiparton cross sections that mimics the power-law
evolution of $Q_\text{sat}$ characteristic of gluon saturation
models~\cite{McLerran:2013una}. In addition, the latter saturation models
consistently connect $Q_\text{sat}$ to the impact parameter of the hadronic
collision, thereby providing a natural dependence of $\langle\pt\rangle$ on the
final particle multiplicity in the event.

\section{Summary}

\label{sec:summary}

Transverse momentum spectra have been measured for different charged hadron
species produced in inelastic pp collisions at $\sqrt{s} = 13\TeV$. Charged
pions, kaons, and protons are identified from the energy deposited in the
silicon tracker and the reconstructed particle trajectory.
The yields of such hadrons at rapidities $\abs{y} < 1$ are studied as a
function of the event charged particle multiplicity measured in the
pseudorapidity range $\abs{\eta}<2.4$.
The transverse momentum (\pt) spectra are well described by fits using the
Tsallis-Pareto parametrization. The ratios of the yields of oppositely charged
particles are close to unity, as expected in the central rapidity region for
collisions at this center-of-mass energy.
The average \pt is found to increase with particle mass and event
multiplicity, and shows features a slow (logarithmiclike or power-law)
dependence on $\sqrt{s}$.

As observed in lower-energy data, the $\langle\pt\rangle$ and the
ratios of particle yields are strongly correlated with event particle
multiplicity.
The {\PYTHIA}8 CUETP8M1 event generator reproduces most features of the
measured distributions, which represents a success of the preceding tuning of
this model, and \EPOSLHC\ also gives a satisfactory description of several
aspects of the data.
Although soft QCD effects are intertwined with other effects, the present
results could be used to further constrain models of hadron production and to
contribute to a better understanding of multiparton interactions, parton
hadronization, and final-state effects in high-energy hadron collisions.

\begin{acknowledgments}
We congratulate our colleagues in the CERN accelerator departments for the excellent performance of the LHC and thank the technical and administrative staffs at CERN and at other CMS institutes for their contributions to the success of the CMS effort. In addition, we gratefully acknowledge the computing centers and personnel of the Worldwide LHC Computing Grid for delivering so effectively the computing infrastructure essential to our analyses. Finally, we acknowledge the enduring support for the construction and operation of the LHC and the CMS detector provided by the following funding agencies: BMWFW and FWF (Austria); FNRS and FWO (Belgium); CNPq, CAPES, FAPERJ, and FAPESP (Brazil); MES (Bulgaria); CERN; CAS, MoST, and NSFC (China); COLCIENCIAS (Colombia); MSES and CSF (Croatia); RPF (Cyprus); SENESCYT (Ecuador); MoER, ERC IUT, and ERDF (Estonia); Academy of Finland, MEC, and HIP (Finland); CEA and CNRS/IN2P3 (France); BMBF, DFG, and HGF (Germany); GSRT (Greece); OTKA and NIH (Hungary); DAE and DST (India); IPM (Iran); SFI (Ireland); INFN (Italy); MSIP and NRF (Republic of Korea); LAS (Lithuania); MOE and UM (Malaysia); BUAP, CINVESTAV, CONACYT, LNS, SEP, and UASLP-FAI (Mexico); MBIE (New Zealand); PAEC (Pakistan); MSHE and NSC (Poland); FCT (Portugal); JINR (Dubna); MON, RosAtom, RAS, RFBR and RAEP (Russia); MESTD (Serbia); SEIDI, CPAN, PCTI and FEDER (Spain); Swiss Funding Agencies (Switzerland); MST (Taipei); ThEPCenter, IPST, STAR, and NSTDA (Thailand); TUBITAK and TAEK (Turkey); NASU and SFFR (Ukraine); STFC (United Kingdom); DOE and NSF (USA).

\hyphenation{Rachada-pisek} Individuals have received support from the Marie-Curie program and the European Research Council and Horizon 2020 Grant, contract No. 675440 (European Union); the Leventis Foundation; the A. P. Sloan Foundation; the Alexander von Humboldt Foundation; the Belgian Federal Science Policy Office; the Fonds pour la Formation \`a la Recherche dans l'Industrie et dans l'Agriculture (FRIA-Belgium); the Agentschap voor Innovatie door Wetenschap en Technologie (IWT-Belgium); the Ministry of Education, Youth and Sports (MEYS) of the Czech Republic; the Council of Science and Industrial Research, India; the HOMING PLUS program of the Foundation for Polish Science, cofinanced from European Union, Regional Development Fund, the Mobility Plus program of the Ministry of Science and Higher Education, the National Science Center (Poland), contracts Harmonia 2014/14/M/ST2/00428, Opus 2014/13/B/ST2/02543, 2014/15/B/ST2/03998, and 2015/19/B/ST2/02861, Sonata-bis 2012/07/E/ST2/01406; the National Priorities Research Program by Qatar National Research Fund; the Programa Clar\'in-COFUND del Principado de Asturias; the Thalis and Aristeia programs cofinanced by EU-ESF and the Greek NSRF; the Rachadapisek Sompot Fund for Postdoctoral Fellowship, Chulalongkorn University and the Chulalongkorn Academic into Its 2nd Century Project Advancement Project (Thailand); and the Welch Foundation, contract C-1845.
\end{acknowledgments}

\clearpage
\bibliography{auto_generated}

\cleardoublepage \appendix\section{The CMS Collaboration \label{app:collab}}\begin{sloppypar}\hyphenpenalty=5000\widowpenalty=500\clubpenalty=5000\textbf{Yerevan Physics Institute,  Yerevan,  Armenia}\\*[0pt]
A.M.~Sirunyan, A.~Tumasyan
\vskip\cmsinstskip
\textbf{Institut f\"{u}r Hochenergiephysik,  Wien,  Austria}\\*[0pt]
W.~Adam, E.~Asilar, T.~Bergauer, J.~Brandstetter, E.~Brondolin, M.~Dragicevic, J.~Er\"{o}, M.~Flechl, M.~Friedl, R.~Fr\"{u}hwirth\cmsAuthorMark{1}, V.M.~Ghete, C.~Hartl, N.~H\"{o}rmann, J.~Hrubec, M.~Jeitler\cmsAuthorMark{1}, A.~K\"{o}nig, I.~Kr\"{a}tschmer, D.~Liko, T.~Matsushita, I.~Mikulec, D.~Rabady, N.~Rad, B.~Rahbaran, H.~Rohringer, J.~Schieck\cmsAuthorMark{1}, J.~Strauss, W.~Waltenberger, C.-E.~Wulz\cmsAuthorMark{1}
% ...
\vskip\cmsinstskip
% ...
1:~~Also at Vienna University of Technology, Vienna, Austria\\
% ...

\end{sloppypar}
\end{document}